\documentclass[sigconf]{acmart}

\usepackage{algorithm}

\usepackage{graphicx} 
\usepackage{graphics}
\usepackage{xspace}
\usepackage{makecell}
\usepackage{colortbl}
\usepackage{enumitem}
\usepackage{pifont}
\usepackage{amsfonts}
\usepackage{float}
\usepackage{subfig}
\usepackage{fontawesome5}
\usepackage{multirow}
\usepackage{subfig}
\usepackage{textcomp}
\usepackage{stfloats}
\usepackage{url}
\usepackage{verbatim}

\newcommand{\appname}{{\sc Attain}\xspace}
\newcommand{\vszzname}{{\sc V-SZZ}\xspace}
\newcommand{\llmname}{{\sc LLM4SZZ}\xspace}
\newcommand{\pocname}{{\sc Exploit}\xspace}

\definecolor{myyellow}{HTML}{FFF2CC}
\definecolor{myblue}{RGB}{255,255,255}

\definecolor{lightgray}{gray}{0.95}

\AtBeginDocument{%
  }

\author{Xinwei Mao}
\orcid{}
\affiliation{%
  \institution{School of Software Technology, Zhejiang University}
   \city{Ningbo}
  \country{China}
}
\email{xw_mao@zju.edu.cn}

\author{Zirui Chen}
\orcid{0009-0004-6236-9150}
\affiliation{%
  \institution{The State Key Laboratory of Blockchain and Data Security, Zhejiang University}
   \city{Hangzhou}
  \country{China}
}
\email{chenzirui@zju.edu.cn}

\author{Xing Hu}
\orcid{0000-0003-0093-3292}
\authornote{Corresponding authors}
\affiliation{
  \institution{The State Key Laboratory of Blockchain and Data Security, Zhejiang University}
   \city{Hangzhou}
  \country{China}
}
\email{xinghu@zju.edu.cn}

\author{Xin Xia}
\orcid{0000-0002-6302-3256}
\affiliation{
  \institution{The State Key Laboratory of Blockchain and Data Security, Zhejiang University}
   \city{Hangzhou}
  \country{China}
}
\email{xin.xia@acm.org}

\begin{document}

\title[ATTAIN: Automated Exploit Failure Analysis through Trace-Driven Diff Analysis]{ATTAIN: Automated Exploit Failure Analysis \\ through Trace-Driven Diff Analysis}

\begin{abstract}


Exploits are widely used to check whether library vulnerabilities appear in different versions and to mark affected version ranges. Exploit-based checks sometimes fail because exploits stop running on many versions after API or environment changes. Commit-based methods, such as SZZ-style analysis, sometimes miss the right introduce commits and spread labels incorrectly along long version chains. These problems leave many affected versions unlabeled or wrongly labeled and make manual exploit failure analysis very expensive and impractical at scale.

We present \appname, a trace-driven diff analysis framework with three modules to assess vulnerability presence across evolving library versions. The modules are trace construction, diff exploration, and affected-version judgment. The trace construction module executes an exploit across historical library versions and compares their behaviors to capture cross-version execution divergences. Using these divergences, the diff exploration module guides an LLM through a finite-state tool loop to autonomously search over version changes and collect vulnerability-relevant diff hunks. The affected-version judgment module reasons over the collected evidence to determine whether the vulnerability exists in each version and outputs the affected version range.

We evaluate \appname on an extensive dataset comprising 224 CVEs and 25,943 library versions across 128 libraries. \appname achieves an F1-score of 93.24\%, outperforming the commit-based methods \vszzname and \llmname by 116.28\% and 33.30\% respectively. \appname uses short tool-guided prompts and a fixed number of iterations, keeping token usage low. It matches or surpasses existing methods on frequent CWE types, including cases where exploit runs fail for non-vulnerability reasons or commit messages do not clearly delimit affected versions.

\end{abstract}

\ccsdesc[500]{Security and privacy~Software security engineering}

\keywords{Library Vulnerabilities, Exploit Trace, Affected Version}

\maketitle

\section{Introduction}
Vulnerabilities in open-source libraries have become a critical security concern in modern software development, because these libraries are widely reused to reduce redundant implementation effort and accelerate software delivery~\cite{Wu2024Vision,Markus1,Synopsys1,Kula1,Zhan2025React}, allowing vulnerable upstream components to affect numerous downstream projects~\cite{Zhou2024Magneto,zirui2024exploiting,He2023Dependent,shuhan2025vul,Zhan2024PS3,Pan2024Assessment,Pan2023Type,Pan2022Reports,jiayuan2022fix,jiayuan2023cole,Bavota1,kula2018developers,Li2023downstream,chen2025generatingmitigationsdownstreamprojects}. To assess these risks, downstream maintainers need to determine whether their projects depend on vulnerable library versions~\cite{chen2026largescaleempiricalstudygeneralizability, chen2025diffploitfacilitatingcrossversionexploit}.

A commonly adopted practice is to consult the affected version ranges disclosed by vulnerability databases, such as NVD and Snyk. However, recent studies have shown that these databases often suffer from inaccuracies and inconsistencies~\cite{Croft2023report,Dong2019Version,Jo2021Report}. To address this challenge, researchers have proposed patch-based approaches~\cite{Wu2024Vision, Bao2022Vszz} that locate vulnerability-related information from security patches and then assess whether these statements or their surrounding structures exist in historical versions.

However, these techniques still face challenges when the introducing commit and the fixing patch are structurally distant. For example, CVE-2023-51080 in Hutool is introduced in version 5.8.22, where the commit~\cite{introduce51080} adds a recursive fallback path in \textit{toBigDecimal}. When receiving malformed numeric inputs, such as `NaN', this recursive path repeatedly invokes number parsing and eventually causes a `StackOverflowError'. The vulnerability is fixed in version 5.8.25 by adding an assertion to reject invalid parsed numbers. Existing patch-based approaches, such as Vision~\cite{Wu2024Vision}, treat all versions before 5.8.25 as affected due to failing to identify the actual vulnerability-introducing change.

To address the above challenge, we need to accurately identify vulnerability-introducing diff hunks beyond structural similarity to the fixing patch. We make two key observations: \ding{182} vulnerability exploit execution traces across versions provide useful signals for localizing related functions; and \ding{183} Deep learning, especially large language models (LLMs), which are widely used due to their capabilities in various code-related tasks~\cite{grotov2024untanglingknotsleveragingllm,  zou2024docbenchbenchmarkevaluatingllmbased, Saboor2025repair, li2022poisonattackdefensedeep, tyen2024llmsreasoningerrorscorrect, Xue2024LLM, Xue2025LLM, yu2025selfadmittedcodegeneratedlarge, gao2026depradaragenticcoordinationcontext, Gao2025LLM}, can reason about whether a change may introduce vulnerability-related behavior. Based on these observations, we propose \appname, an LLM-guided framework that autonomously searches cross-version changes based on exploit execution behavior, collects relevant diff hunks, and determines affected versions based on the collected evidence.

\appname consists of three modules. Given a public exploit and historical library versions, the \textbf{trace construction module} executes a public exploit across historical library versions and compares their behaviors to capture cross-version execution divergences if the exploit triggers the vulnerability in one version but fails to reproduce or compile in the nearby version. Second, for each version where the exploit is not reproduced, the \textbf{diff exploration module} uses these divergences to guide an autonomous search over version changes and collect vulnerability-relevant diff hunks. Third, the \textbf{affected-version judgment module} reasons over the collected evidence to determine whether the vulnerability exists in each version and outputs the affected version range.

We evaluate \appname on an extensive dataset comprising 224 CVEs and 25,943 library versions across 128 Java libraries. \appname achieves an F1-score of 93.24\%, outperforming the commit-based methods \vszzname and \llmname by 116.28\% and 33.30\% respectively. Meanwhile, \appname uses short trace-guided prompts and bounded iterations to keep token costs low enough for practical deployment. Our ablation and CWE-level analyses show that trace-diff guidance, dynamic diff search, rule-based aggregation, and version-chain backfill all contribute to effectiveness and coverage. These components are particularly helpful when exploits fail for non-vulnerability reasons or commit messages do not clearly delimit affected versions and traces.

This paper makes the following main contributions:
\begin{itemize}[leftmargin=*]
    \item We propose \appname, a trace-driven diff analysis framework that autonomously searches cross-version changes from exploit execution traces, collects vulnerability-relevant diff hunks, and determines affected versions from the collected evidence. The source code and dataset of \appname are publicly available in the online replication package~\cite{replication}.
    \item We conduct an evaluation on 224 CVEs and 25,943 library versions across 128 libraries. \appname achieves an F1-score of 93.24\%, outperforming the commit-based methods \vszzname and \llmname by 116.28\% and 33.30\% respectively.
\end{itemize}

The remainder of this paper is organized as follows.
Section~\ref{sec:motivation} motivates our work.
Section~\ref{sec:approach} presents \appname's three modules.
Section~\ref{sec:setup} describes the experimental setup.
Section~\ref{sec:evaluation} reports results against baselines, with ablation and per-CWE analyses.
Section~\ref{sec:discussion} discusses typical failures and threats.
Section~\ref{sec:related} reviews related work.
Section~\ref{sec:conclusion} summarizes the paper.

\section{Motivation}
\label{sec:motivation}

This section explains why identifying affected versions is challenging. We first show that patch-based approaches are inaccurate when the introducing and fixing commits are structurally distant, motivating the use of exploits. We then discuss why failed exploit runs are ambiguous and lead to low recall. Finally, we outline how \appname analyzes exploit failures along a version chain to strengthen affected-version identification.

\subsection{Exploit Signals}

Existing patch-based approaches locate vulnerability-related information from security patches and check its presence in historical versions~\cite{Wu2024Vision}.
However, they become inaccurate when the vulnerability-introducing commit and the fixing patch are structurally distant.
For example, CVE-2023-51080 in Hutool introduces a recursive fallback path in version~5.8.22 and fixes it with an assertion in 5.8.25. 

In contrast, running the exploit on 5.8.22 directly triggers the vulnerability, providing unambiguous evidence.
This illustrates that successful exploit executions yield high-precision affected-version signals.This example also clarifies why we choose exploit-driven analysis as the core signal.
For CVE-2023-51080, patch-only matching over-approximates affected versions because the introducing and fixing commits are structurally distant, while exploit execution directly tests vulnerability semantics on each concrete version.
Moreover, exploit outcomes remain useful even when they are negative: with cross-version traces and diffs, a failed run can be diagnosed as pre-introduction, true fix, or exploit-version mismatch, turning ambiguous negatives into actionable evidence and improving recall without sacrificing precision.

However, a failed exploit run is ambiguous: it may mean the vulnerability has not yet been introduced, has already been fixed, or remains present but the exploit fails due to breaking changes.
Existing workflows rarely distinguish these cases, leading to low recall or requiring extensive manual analysis.

\subsection{Motivating Examples}

Our key observation is that exploit failures do not occur in isolation: they happen along a version chain where code and behavior evolve over time.
\appname combines cross-version execution traces with version diffs to explain exploit failures at the level of concrete code changes, aiming to separate two categories: (1)~\emph{pre-introduction failures}, where the vulnerable behavior has not yet been introduced; and (2)~\emph{breaking-change failures}, where the vulnerability persists but API or environment changes prevent the exploit from running.
We next discuss one representative case for each category.

\subsubsection{Case 1: Pre-introduction}

CVE-2023-1436 in jettison affects version~1.3.1, and our exploit explicitly targets the constructor \texttt{JSONObject(Map map)}.
The exploit builds a self-referential input map and invokes JSON object construction; inside the constructor, nested maps are recursively wrapped (e.g., \texttt{myHashMap.put(k, new JSONObject((Map) v));}), which repeatedly re-enters the same object graph.
As a result, the exploit triggers an infinite recursion and eventually a StackOverflowError in version~1.3.1.

\begin{figure}[t]
  \centering
  \includegraphics[width=\linewidth]{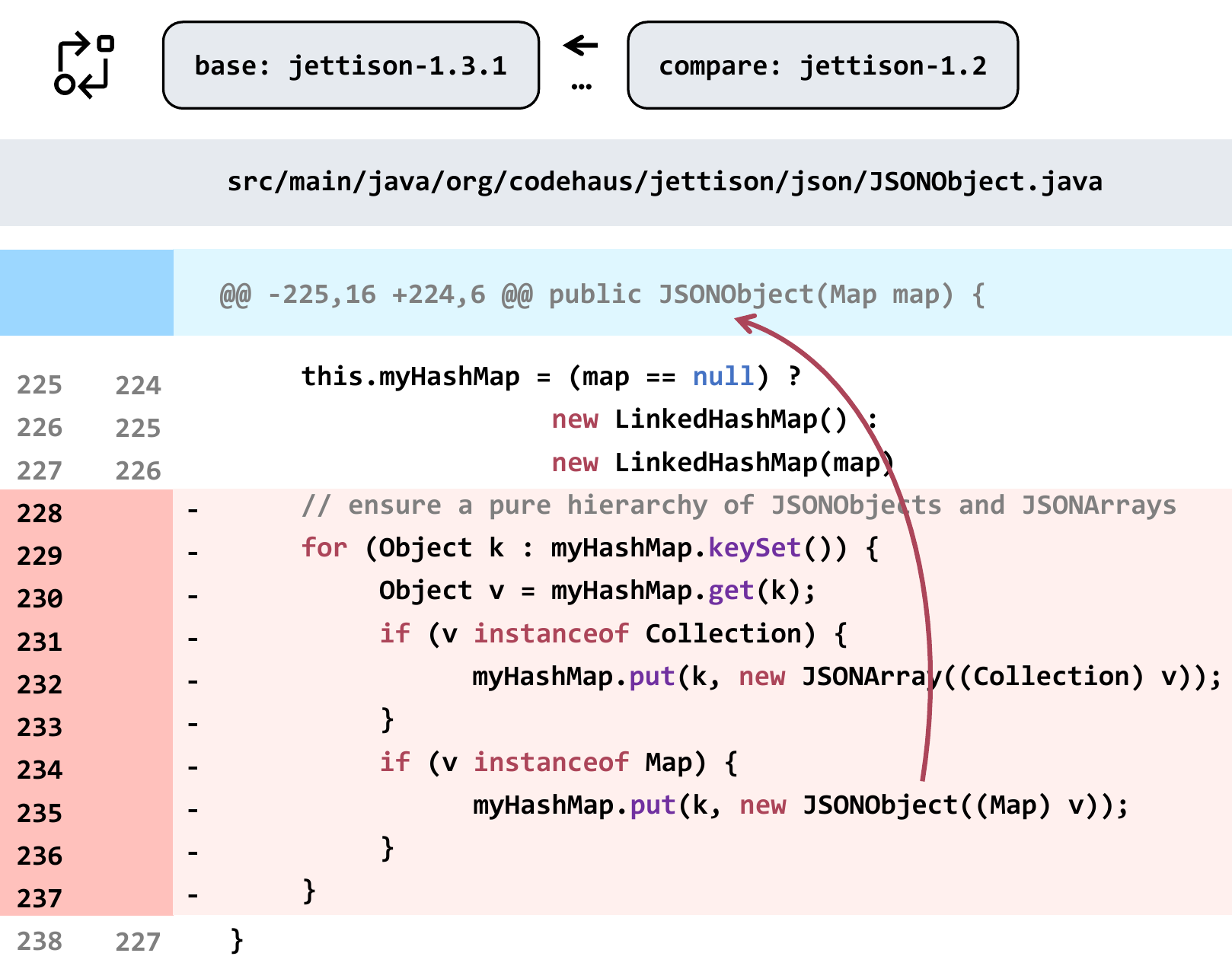}
  \caption{Pre-introduction failure case for CVE-2023-1436 in org.codehaus.jettison:jettison, comparing versions 1.3.1 (reference, affected) and 1.2 (target, not affected).}
  \label{fig:case1-pre-introduction}
\end{figure}

Running the same exploit on version~1.2 does not reproduce the stack overflow.
As shown in Figure~\ref{fig:case1-pre-introduction}, \appname analyzes the cross-version traces and diffs: in the reference trace (1.3.1), the execution reaches the recursive path that causes the stack overflow; in the target trace (1.2), the execution diverges earlier and never enters this path, and the corresponding diff shows that the vulnerable recursive logic is absent.
Based on this combination, \appname classifies 1.2 as a \emph{pre-introduction failure}: the exploit fails because the vulnerable behavior has not yet been introduced.

\subsubsection{Case 2: Breaking Change}

CVE-2021-21351 in xstream is a deserialization vulnerability where a crafted XML payload drives the reflective deserialization pipeline of XStream, potentially leading to remote code execution.
Both versions 1.4.6 and 1.4.5 are affected.

\begin{figure}[t]
  \centering
  \includegraphics[width=\linewidth]{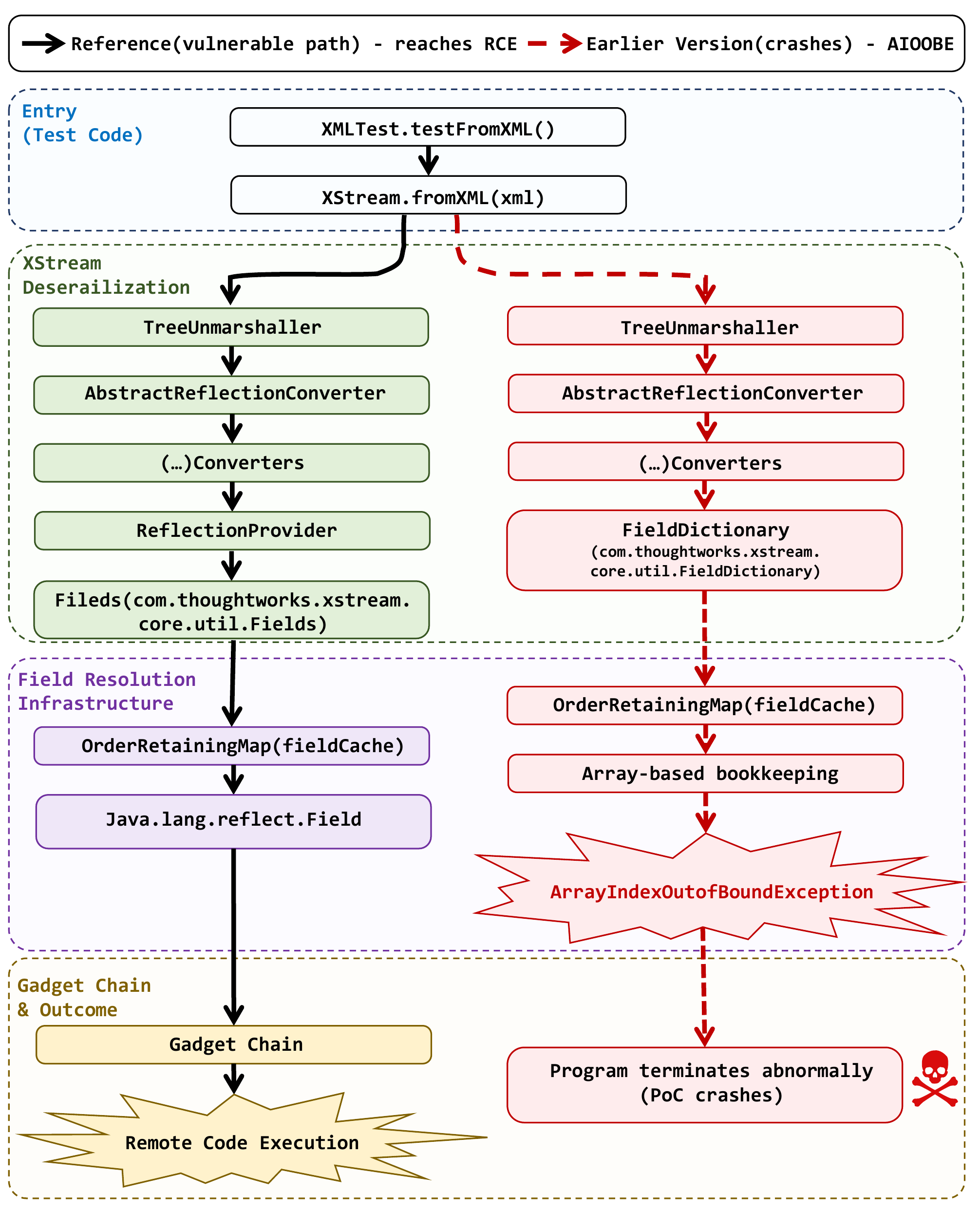}
  \caption{Breaking-change failure case for CVE-2021-21351 in com.thoughtworks.xstream:xstream, comparing versions 1.4.6 (reference, affected) and 1.4.5 (target, affected).}
  \label{fig:case2-breaking-change}
\end{figure}

In version~1.4.6 (reference), the exploit follows the solid path in Figure~\ref{fig:case2-breaking-change}, traversing the reflective deserialization pipeline and reaching the RCE-enabling gadget chain.
In version~1.4.5, the exploit crashes early with \texttt{ArrayIndexOutOfBoundsException} inside \texttt{OrderRetainingMap.entrySet}, before reaching the vulnerable reflection logic.
To distinguish this from a true non-affected case, \appname again uses a multi-signal check.
First, the failure signature is a runtime crash in field-resolution infrastructure (AIOOBE), not a clean disappearance of vulnerable semantics.
Second, trace alignment shows that both versions share the same upper reflective pipeline (\texttt{TreeUnmarshaller} and \texttt{AbstractReflectionConverter}) and diverge around \texttt{FieldDictionary}/\texttt{OrderRetainingMap}.
Third, the version diff highlights behavior changes in helper code such as \texttt{core.util.Fields}, while the core vulnerable deserialization chain is still present.
Together, these signals indicate \emph{exploit invalidation due to breaking changes}: version~1.4.5 remains affected, but this exploit instance no longer runs to completion on that environment.




\section{Proposed Approach}
\label{sec:approach}

\begin{figure*}[t]
  \centering
  \includegraphics[width=\textwidth]{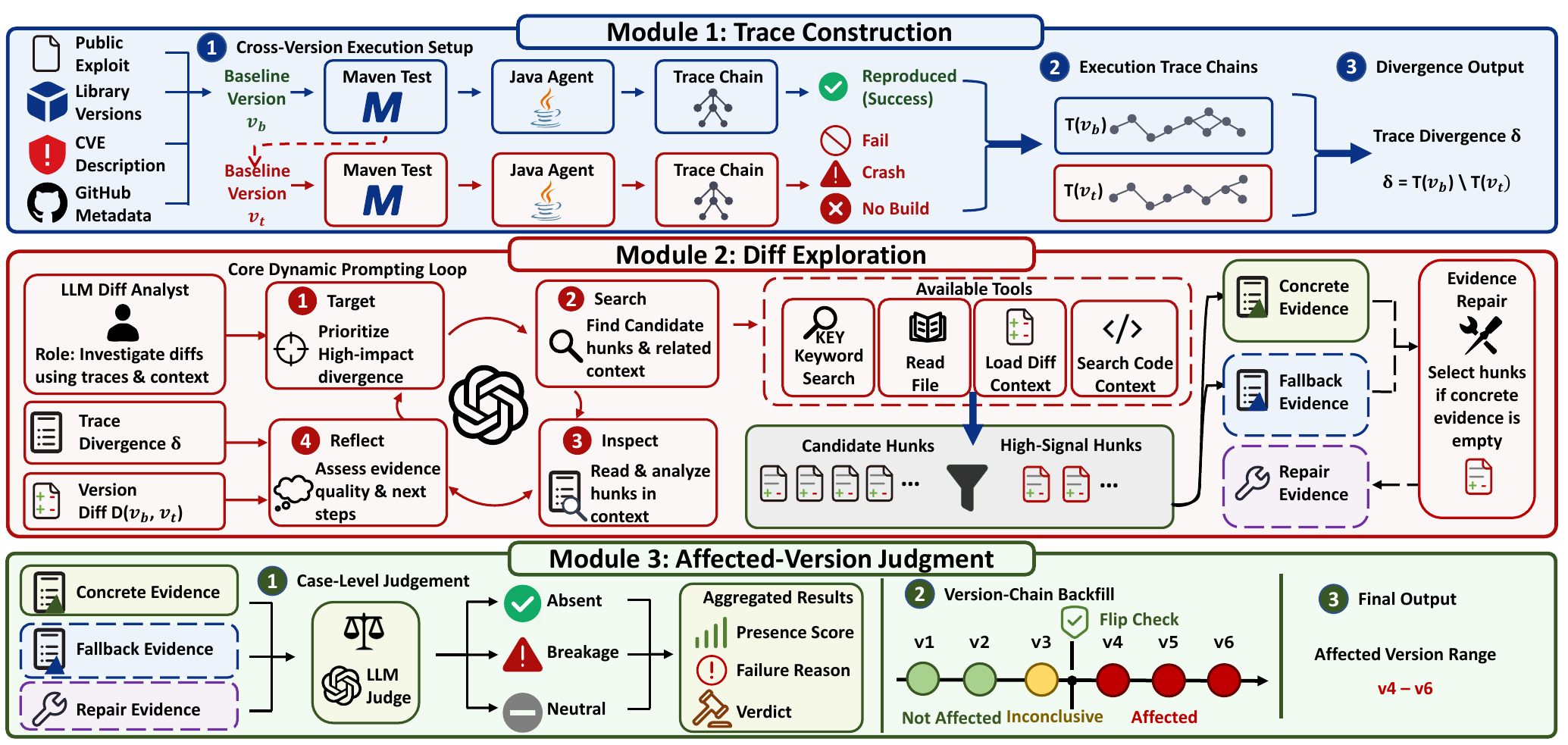}
  \caption{The overall framework of \appname.}
  \label{fig:flowchart}
\end{figure*}

We now describe \appname, our approach to cross-version affected-version determination. The method operates through three modules: the \textbf{Trace Construction Module}, which captures cross-version execution divergences; the \textbf{Diff Exploration Module}, which recovers vulnerability-relevant evidence from version diffs; and the \textbf{Affected-Version Judgment Module}, which reasons over the evidence and extends labels along the version chain. Figure~\ref{fig:flowchart} presents the overall workflow.

\subsection{Trace Construction Module}

Given a public exploit and the historical versions of a library, the trace construction module executes the exploit across versions and compares their behaviors. Its goal is to capture cross-version execution divergences when the exploit triggers the vulnerability in one version but fails to reproduce or compile in a nearby version. When multiple exploits exist for the same CVE, each exploit runs independently through the entire pipeline.

We start from a known CVE and its publicly available exploit. We also have two versions of the same library. One is the reference version $v_b$, where the exploit can trigger the vulnerability. The other is the target version $v_t$, where the exploit behaves differently. \appname runs the exploit on both versions and records execution traces. These traces form an execution trace chain.

For each version, \appname uses the Maven testing framework to run the exploit. A lightweight Java agent based on ASM bytecode manipulation instruments every non-abstract method within the target library's package namespace, recording fully qualified class name, method name, and descriptor on each method entry. The execution trace chain includes three parts:

\begin{itemize}[leftmargin=*]
    \item Method execution sequence: the method call path during exploit execution, shown as fully qualified class and method names.
    \item Dependency tree: the complete dependency graph obtained from the build system.
    \item Execution observations: whether compilation or runtime succeeds or fails, together with exception types and assertion failure messages.
\end{itemize}

Trace divergence. Let $T(v)$ denote the method enter sequence obtained by running the exploit on version $v$. A trace divergence $\delta$ is computed by element-wise comparison of $T(v_b)$ and $T(v_t)$: the system identifies the first index where the two sequences differ, and also records the set of methods unique to each version. Formally:
$$\delta = T(v_b) \setminus T(v_t)$$
A trace divergence can appear in multiple forms: a compilation failure, a runtime exception, a difference in the method call sequences, or an early termination of one trace relative to the other.

The two trace chains may differ from each other. This difference directly shows behavioral differences between the two versions. If the exploit triggers the vulnerability on $v_b$ but behaves differently on $v_t$, then the vulnerability trigger path has diverged across the two versions. Recursive calls naturally produce repeated entries in the trace and are handled by the element-wise comparison without special treatment. For extremely large traces (exceeding 1~GB), the system switches to a lightweight mode that infers divergence from execution observations and trace file sizes rather than full element comparison.

Version pair preprocessing. Before running the trace comparison, \appname needs to decide which version pairs to analyze and which commits are the relevant patches. The first half of this pipeline resolves version boundaries. It starts from the ground truth execution results. The system finds the upstream GitHub repository for each CVE, then sorts all tested versions by Maven release timestamp and constructs a branch-aware version tree. A version is placed on the main release chain unless its timestamp regresses relative to its predecessor, in which case it is attached to the corresponding maintenance branch. Adjacent version pairs are defined as consecutive versions along this tree where the exploit outcome flips from triggering the vulnerability to not triggering it, or vice versa. Versions not present in the execution results are excluded; unresolved versions are concatenated to the nearest resolved predecessor. Across the 224 CVEs, this algorithm generates 845 boundary version pairs, of which 459 unique pairs are selected for LLM-based analysis. Maven version strings do not always match GitHub tag names. The system queries the GitHub Tags API for each repository and matches Maven versions to GitHub tags by prefix completion, dot-to-underscore conversion, and strict pattern matching. For each version pair, the system calls the GitHub Compare API to get the list of commits between the two tags.

The second half of the pipeline filters commits and builds patch context. Given the commit list and the CVE description, the system uses dual-encoder retrieval to rank commit messages by their similarity to the CVE description. The top-$k$ candidates are then passed to an LLM, which selects the most likely fix commits. Next, the system fetches the full commit details from GitHub, filtering out non-code files such as documentation, configuration, test files, and build scripts. Only commits with remaining valid files are kept. For each remaining commit, the system parses the diff hunks and fetches the surrounding source lines from GitHub. It applies two levels of pruning. Line-level pruning removes comments, import statements, and logging calls. Hunk-level pruning discards test files and empty diffs. Finally, a bi-encoder model scores each file-level diff against the CVE description. Files scoring above a threshold are labeled as patch. If no file passes the threshold, an LLM checks all candidates in descending score order and stops at the first positive verdict, yielding the final set of patch diffs.

\subsection{Diff Exploration Module}

For each version where the exploit is not reproduced, the diff exploration module uses the observed divergences to guide an autonomous search over version changes and collect vulnerability-relevant diff hunks. The divergence is passed to the LLM as a structured summary containing the first point of trace mismatch (with the diverging method signatures), the sets of methods unique to each version, and the execution observations. In a multi-exploit setting, each exploit independently feeds its own divergence into this module.

Version diff. Let $D(v_b, v_t)$ denote the set of all diff hunks between the source code or bytecode of $v_b$ and $v_t$. Each diff hunk $h \in D$ is a localized code change. It includes the file path, the changed lines, and the surrounding context.

Trace-diff context. A trace-diff context $C_i$ links an observed $\delta_i$ between $v_b$ and $v_t$ to its search state $s_i$ and the collected evidence hunks. It is defined as:
$$C_i = (\delta_i,\; s_i,\; E_i^{\text{concrete}},\; E_i^{\text{fallback}})$$
Here $E_i^{\text{concrete}}$ contains diff hunks that directly explain $\delta_i$. $E_i^{\text{fallback}}$ contains supplementary evidence such as trace-diff summaries, dependency-tree diffs, and build configuration diffs. These files provide contextual information but cannot by themselves prove that the vulnerability is absent.

Evidence repair. When the collected evidence $E_i^{\text{concrete}}$ is empty, the system triggers the evidence repair procedure. This procedure selects the highest-scoring candidate hunks from $D(v_b, v_t)$ based on term matching against the scenario keywords and previously observed symbols. The repair result is denoted as:
$$E_i^{\text{repair}} = \text{top-}k \left( \text{score}(h,\; \delta_i) \mid h \in D \right)$$
where $\text{score}(h, \delta_i)$ measures the relevance between hunk $h$ and $\delta_i$.

\begin{figure}[t]
  \centering
  \includegraphics[width=\linewidth]{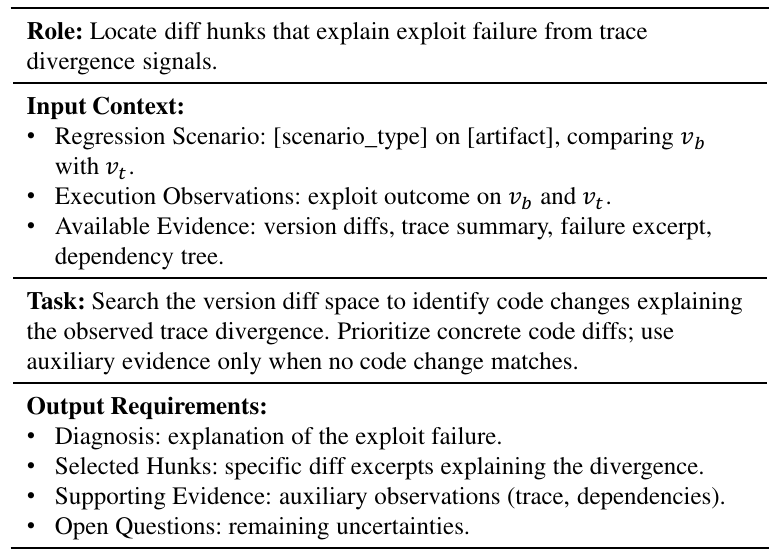}
  \caption{The Prompt for Diff Exploration}
  \label{fig:prompt-1}
\end{figure}

When a trace divergence $\delta_i$ is detected, the system constructs the corresponding $C_i$ and searches $D(v_b, v_t)$ for the evidence hunks associated with $C_i$. Since $D(v_b, v_t)$ may contain many candidate hunks, directly recovering the evidence that explains $\delta_i$ is difficult. We therefore use a tool-augmented dynamic prompting approach (Figure~\ref{fig:prompt-1}) to recover $E_i^{\text{concrete}}$ for $C_i$; when direct evidence is unavailable, the procedure falls back to $E_i^{\text{fallback}}$ and may subsequently derive $E_i^{\text{repair}}$.


\subsubsection{Role.}
The LLM plays the role of a dependency version regression analyst. Its core task is to recover $E_i^{\text{concrete}}$ from $D(v_b, v_t)$ for each $C_i$. These retained hunks form a minimal set of high-signal inputs for the next step of vulnerability presence judgment.

The specific responsibilities are:
\begin{itemize}[leftmargin=*]
    \item Understanding the behavioral signal represented by $\delta_i$ and the associated search state $s_i$ in $C_i$, including compilation failure and runtime exceptions.
    \item Locating candidate hunks $h \in D$ and organizing the retained evidence as $E_i^{\text{concrete}}$ or $E_i^{\text{fallback}}$.
    \item Telling apart evidence that the vulnerability is absent from exploit breakage evidence within the retained evidence set. Evidence that the vulnerability is absent means changes that remove or block the vulnerable path. Exploit breakage evidence means changes that only break the exploit harness but do not affect whether the vulnerability is still present.
\end{itemize}

\subsubsection{Goals.}
The goals of the dynamic prompting stage are as follows.
\begin{itemize}[leftmargin=*]
    \item Recover $E_i^{\text{concrete}}$ for each $C_i$ by identifying the hunks in $D(v_b, v_t)$ that best explain $\delta_i$.
    \item Look at API-level changes first, including method signature changes and class removals.
    \item When API diffs are too shallow and contain only class signatures, field descriptors, or other declaration-only changes, go deeper into bytecode-level diffs to see the method-body context.
    \item When no direct hunk can be retained in $E_i^{\text{concrete}}$, allow citing supplementary evidence in $E_i^{\text{fallback}}$. This evidence must be clearly marked as fallback evidence and cannot alone prove that the vulnerability is absent.
\end{itemize}

\subsubsection{Available tools.}
The LLM can use the following capabilities to recover evidence from $D(v_b, v_t)$.

\begin{itemize}[leftmargin=*]
    \item{Keyword search.} This capability narrows the candidate space by retrieving file-line pairs associated with keywords derived from $\delta_i$ and $s_i$. It returns a ranked list of candidates that helps identify promising hunks $h \in D$.
    
    \item{Read file.} This capability inspects the local content of a candidate hunk $h$ within a narrow window, allowing the model to verify whether the candidate should be retained in $E_i^{\text{concrete}}$.
    
    \item{Load diff context.} This capability resolves a candidate diff hunk $h$ to its surrounding reference and target code context, enabling the model to assess whether the change belongs to $E_i^{\text{concrete}}$.
    
    \item{Search code context.} This capability recovers broader program structure relevant to $C_i$, such as class-level or method-level context, when the local diff fragment is insufficient.
\end{itemize}

\subsubsection{State.}
The core of tool-augmented dynamic prompting is a finite-state iterative loop. At each round, the LLM decides its next action based on the current state. The state transitions work over the current $C_i$ as follows.

\begin{itemize}[leftmargin=*]
    \item{Determine collection target.}
    The input is the scenario type, failure indicators, and trace keywords. The LLM reads the overview of the version diff and the failure symptoms, then initializes the search state $s_i$ in $C_i$ with target classes, methods, or exception symbols as starting points.

    \item{Construct search query.}
    The LLM uses the current $s_i$ together with $\delta_i$-based keywords and failure-excerpt keywords. It then performs keyword-based retrieval over the analysis workspace to identify candidate hunks $h \in D$, producing a ranked list of candidate diff paths by relevance score.

    \item{Reflect.}
    The LLM examines the current $C_i$, including the retrieved candidates and any retained evidence. It makes three checks: Is the current evidence enough to explain $\delta_i$? If yes, go to the next step. Are the API diffs too shallow and only show declarations? If so, it retrieves expanded diff context to refine $E_i^{\text{concrete}}$ or gathers broader code context to recover the surrounding program structure. Are some important classes missing? If so, update $s_i$ and go back to construct search query.
\end{itemize}

During reflection, the LLM must respect the following constraints:
\begin{itemize}[leftmargin=*]
    \item The output must contain exactly four sections: Diagnosis, Selected Hunks, Supporting Evidence, and Open Questions.
    \item Declaration-only changes in API diffs must not be used as sufficient evidence to finalize the answer.
    \item Constructor, collection, map, and self-reference style vulnerabilities must be expanded to the complete changed method or constructor context.
    \item Lower priority should be given to logger-only diffs, anonymous inner-class diffs, and enum-reordering diffs.
\end{itemize}

Update target and query. If reflection finds that the retained evidence in $C_i$ is not enough, the LLM extracts new symbols such as class names and method names from the current diff content, updates $s_i$, and merges new keywords from $\delta_i$. It then returns to construct search query with the updated $s_i$.

If reflection finds that $E_i^{\text{concrete}}$ is sufficient, the LLM outputs the final answer as a structured report comprising four sections Diagnosis, Selected Hunks, Supporting Evidence, and Open Questions. If the Selected Hunks section has no valid evidence path, the system derives $E_i^{\text{repair}}$. This ensures the output always contains citable evidence for subsequent judgment.

\begin{figure}[t]
  \centering
  \includegraphics[width=\linewidth]{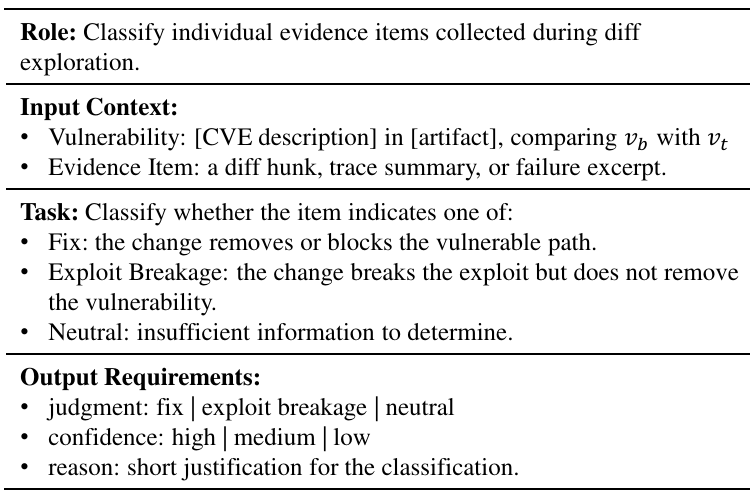}
  \caption{The Prompt for Classfying Evidence}
  \label{fig:prompt-2}
\end{figure}

The iterative loop stops when one of these conditions is met:
\begin{itemize}[leftmargin=*]
    \item The maximum tool-call round budget is reached. The budget changes based on scenario type. Compile failure is limited to one round. Process failure is limited to two rounds. Trace diff allows more rounds for deeper exploration.
    \item The LLM stops calling tools and directly produces the final structured response.
    \item $E_i^{\text{repair}}$ has been derived.
\end{itemize}

Once $E_i^{\text{concrete}}$ or $E_i^{\text{repair}}$ has been derived, the evidence set is passed to the next stage for vulnerability presence judgment.


\subsection{Affected-Version Judgment Module}

Finally, the affected-version judgment module reasons over the collected evidence to determine whether the vulnerability exists in each version and outputs the affected version range. It consists of a case-level vulnerability judgment module and a version-chain backfill module.

\begin{figure}[t]
  \centering
  \includegraphics[width=\linewidth]{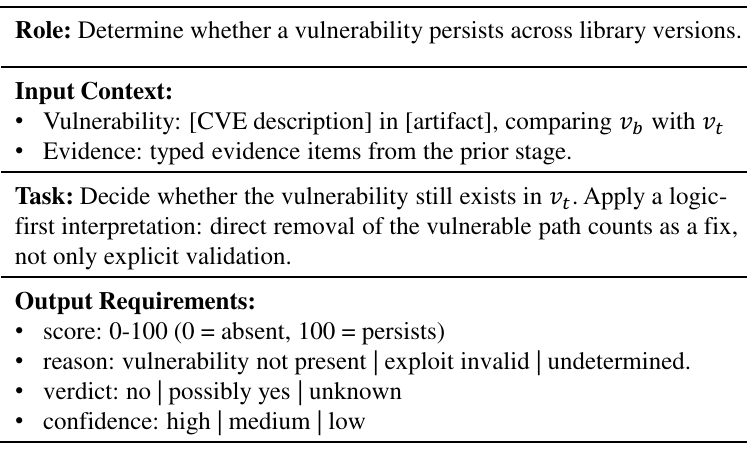}
  \caption{The Prompt for Case-Level Vulnerability Judgment}
  \label{fig:prompt-3}
\end{figure}

\subsubsection{Case-Level Vulnerability Judgment}

After the dynamic prompting stage constructs $C_i$ and recovers its retained evidence, the system needs to judge whether the vulnerability still exists in the target version and why the exploit failed. This process has three modules: single-hunk evidence typing, case-level vulnerability judgment, and rule-based label compression.

Each retained hunk $h$ in $E_i^{\text{concrete}}$ is individually examined by an LLM (Figure~\ref{fig:prompt-2}); when direct evidence is unavailable, the judgment instead relies on the derived set $E_i^{\text{repair}}$. The LLM assigns one of three evidence types: evidence that the vulnerability is absent, meaning the code change removes or blocks the vulnerable path; exploit breakage evidence, meaning the code change breaks the exploit harness but does not imply that the underlying vulnerability is absent; or neutral, meaning the evidence is not sufficient to classify. The LLM also outputs a confidence level and a short reasoning chain for each $h$, which serve as input to the subsequent case-level judgment.

After all retained hunks have been typed, a second LLM call (Figure~\ref{fig:prompt-3}) combines them into a case-level judgment. The LLM receives the typed evidence set together with the corresponding $C_i$. It outputs a vulnerability presence score in $[0, 100]$, a failure reason label (vulnerability not present, exploit invalid due to version change, or inconclusive), and a vulnerability presence verdict (no, possibly yes, or inconclusive). It also outputs separate strength ratings for evidence that the vulnerability is not present and for exploit breakage evidence, giving a complete picture.


The case-level LLM outputs are further compressed into a binary label: affected or not affected. This compression uses a rule-based procedure that considers the vulnerability presence score, the verdict, the scenario type, the failure subtype, and the trace-quality signals encoded in $C_i$. For example, when the trace has an effective difference and the LLM verdict is vulnerability not present, the rule may still change the label to not affected with low confidence if the trace evidence is valid and the first divergence is not noise. This rule module ensures that borderline cases are handled conservatively rather than being forced into a binary decision.
\subsubsection{Version Chain Backfill}
\label{subsec:version-backfill}

The previous stages produce seed labels only for the version pairs that were directly analyzed. \appname extends these labels along the version chain to cover the remaining versions. A version is eligible for label extension only when it has not yet received a label from any earlier stage. Extension is confined within a region, defined as (\textit{CVE}, \text{Library}, \text{Execution outcome}, \text{exploit}). The \textit{exploit} dimension isolates labels across different exploits. If multiple exploits disagree on the same version, \appname adopts a conservative safety-first strategy, defaulting to affected. Before extension starts, if a version has not yet received a label and the exploit can trigger the vulnerability on that version, the system directly assigns the affected label as a safe default.

Within each region, the extension follows the semantics of each label type. Labels associated with the patch commit extend from the patch version toward later versions. Labels associated with the vulnerability introduction commit extend from the introduction version toward earlier versions. Labels that indicate the vulnerability is present extend in both directions along the chain.

The most critical case occurs when an affected label is about to extend across a version boundary where the exploit execution outcome flips. This means the exploit triggers the vulnerability on one version but does not on its adjacent version. Such a flip suggests that the vulnerability behavior changes between the two versions, so simply copying the label may be wrong. In this case, the system invokes an LLM to evaluate whether the vulnerability truly persists on the other side of the flip. The LLM receives the version information and the current judgment details. It returns a confidence score between 0 and 1. If the LLM returns very high confidence that the label should not extend, the system blocks the extension for that version. Otherwise, the extension proceeds. This LLM-based check ensures that label extension does not override a genuine behavior change revealed by the execution outcome flip.

\section{Experimental Setup}
\label{sec:setup}

\textbf{Research Questions.} To evaluate the performance of our approach for cross-version vulnerability presence detection, our experiment aims to answer three research questions:

\begin{itemize}[leftmargin=*]
    \item \textbf{RQ1 (Effectiveness):} How effective is \appname for cross-version vulnerability detection compared to existing approaches?
    \item \textbf{RQ2 (Ablation Study):} How much does each component in \appname contribute to the overall detection performance?
    \item \textbf{RQ3 (CWE-Level Analysis):} How does \appname perform across different CWE vulnerability categories?
\end{itemize}

We address RQ1 to evaluate the overall effectiveness of our method and its advantage over existing approaches. We address RQ2 to measure how much each key component helps. These components include the trace-guided diff exploration module, the evidence and rule modules in vulnerability judgment, and the version chain backfill module. We address RQ3 to check whether \appname keeps its advantage across different CWE categories and to find out where each method has strengths or weaknesses.

\subsection{Dataset}

We evaluate \appname on the largest publicly available Java exploit dataset~\cite{chen2026largescaleempiricalstudygeneralizability}. This dataset contains 259 exploits spanning 224 CVE vulnerabilities across 128 libraries in 41 categories, covering 25,943 library versions (e.g., HTTP clients, XML processors, Object Serialization), with 57 libraries ranking in the top 1,000 Maven artifacts. The vulnerabilities cover 61 CWEs (19 in the CWE Top 25) and carry an average CVSS score of 7.75 (62 Critical, 101 High). Notably, the CWEs in this dataset cover at least one CWE for 76.33\% of all Maven vulnerabilities disclosed up to 2025, and all major weakness Pillars under the CWE Research Concepts view. Each CVE has a corresponding exploit and a set of manually verified affected versions, making the dataset suitable for evaluating cross-version vulnerability presence detection.

To build version pairs for analysis, \appname looks at the execution results for each CVE. It finds adjacent versions where the exploit outcome changes, such as from triggering the exploit successfully to failing to trigger it. These boundary version pairs are the main analysis targets. For each pair, \appname collects execution traces on both the reference version $v_b$ and the target version $v_t$. It also generates version diffs and runs the full pipeline to produce a vulnerability presence judgment.

\subsection{Baselines}

To the best of our knowledge, no prior work has directly addressed the problem of detecting vulnerability presence across Java library versions based on execution traces and version diffs. Existing studies on vulnerability version tracking rely on commit analysis or exploit reproduction alone. We include the following three baselines:

\begin{itemize}[leftmargin=*]
    \item \textbf{\vszzname.} This is a variant of the SZZ algorithm~\cite{sliwerski2005changes} for vulnerability version tracking~\cite{Bao2022Vszz}. \vszzname analyzes version control history to find commits that fix vulnerabilities. It then labels versions based on whether they come before or after the fix commit.

    \item \textbf{\llmname.} This baseline improves the traditional SZZ approach with LLM-based commit analysis~\cite{tang2025llm4szz}. It uses a large language model to decide whether a commit is a vulnerability fix. This replaces the keyword-based matching in traditional SZZ.

    \item \textbf{\pocname.} This baseline uses the exploit execution result directly as the detection label. If the exploit triggers the vulnerability, the version is labeled as \textit{affected}. Otherwise, the version is labeled as \textit{not affected}. This baseline only relies on exploit reproduction. It does not perform any cross-version analysis.
\end{itemize}

\subsection{Evaluation Metrics}

Consistent with prior work, we use precision, recall, and F1-score as the main evaluation metrics. Each method produces a label for each version: \textit{affected} or \textit{not affected}. Versions that a method cannot confidently classify are treated as \textit{not affected} in the main evaluation. This follows the convention that unconfirmed versions are considered safe.

For \appname, fine-grained labels such as affected at the introduction commit and affected at the patch commit are unified into binary labels. A label indicating vulnerability presence is mapped to \textit{affected}; otherwise, \textit{not affected}. The ground truth comes from the dataset~\cite{chen2026largescaleempiricalstudygeneralizability}, which identifies the fixing and inducing commits and manually verifies each version's code. A predicted label is correct if it matches the ground truth.

For the CWE-level analysis in RQ3, we group all CVEs by their primary CWE tag. We take the top 10 most frequent CWE categories. The rest are grouped into an \textit{others} category. For each CWE group and each method, we compute precision, recall, and F1 separately. This produces a comparison across vulnerability types.

\subsection{Implementation Details}

All experiments run on Ubuntu 20.04.6 LTS. We use Java 11 as the runtime, following the same configuration as Wu et al.~\cite{Wu2024Vision}. We run exploits with Apache Maven for dependency resolution. To collect execution traces, we inject a lightweight Java agent at runtime that records method-level traces.

For LLM selection, we use DeepSeek-V3 (snapshot 0324) as the primary model. We choose this model because it is open-source, cost-effective, and has demonstrated strong code understanding capabilities. For comparison, running all 224 cases with a proprietary model such as GPT-5.4 would cost approximately \$114, whereas DeepSeek-V3 completes the same analysis for roughly \$7. This model handles the dynamic prompting module, the single-hunk evidence typing, and the case-level vulnerability judgment respectively. All LLM calls go through the standard OpenAI-compatible API interface with a 180-second timeout. DeepSeek-V3 was used with temperature 0.5/0.2/0.0 across different stages, while context length followed provider defaults and prompt size was controlled via application-level truncation.






\section{Experimental Evaluation}
\label{sec:evaluation}

We evaluate the performance of \appname from three perspectives. First, we assess its effectiveness and compare it with baseline methods on the largest dataset of Java library vulnerability exploits, and analyze its strengths and limitations. Second, we conduct an ablation study to demonstrate the contributions of each component to the overall performance. Finally, we analyze the effectiveness of \appname across different CWE types.

\subsection{Effectiveness}

\subsubsection{Performance.}
We evaluate the effectiveness of \appname on a dataset containing 25,943 library versions, of which 13,109 are vulnerability-affected. As shown in Table~\ref{tab:comparison}, \appname achieves an F1-score of 93.24\%, outperforming all baselines. In particular, \appname identifies 11,741 true positives, which is 699 more than the strongest baseline \pocname (11,042). This corresponds to an improvement of 5.33\% in recall, demonstrating the ability of \appname to detect affected versions that exploit execution alone misses.

\begin{table}[htbp]
  \centering
  \caption{Comparison of Different Approaches}
  \label{tab:comparison}
  \begin{tabular}{lcccccc}
    \toprule
    Tool & TP & FP & FN & Pre. (\%) & Rec.(\%) & F1 (\%) \\
    \midrule
    \vszzname      & 4,482  & 3,200 & 8,627 & 58.34  & 34.19 & 43.11 \\
    \llmname   & 7,815  & 1,419  & 5,294  & 84.63 & 59.62 & 69.95 \\
    \pocname  & 11,042 & 89   & 2,067  & 99.20 & 84.23 & 91.11 \\
    \appname    & 11,741 & 335  & 1,368  & 97.23 & 89.56 & 93.24 \\
    \bottomrule
  \end{tabular}
\end{table}

Compared to \pocname, \appname achieves a relative improvement of 2.34\% in F1-score, demonstrating its robustness in handling cases where the exploit fails for reasons unrelated to the vulnerability. \pocname achieves near-perfect precision (99.20\%) because it only labels a version as \textit{affected} when the exploit actually triggers. However, its recall is limited to 84.23\%: when the exploit fails to run on a target version due to API incompatibility or runtime environment changes, \pocname defaults to \textit{not affected}, missing genuinely affected versions. \appname overcomes this limitation by analyzing the root cause of exploit failure through trace-diff exploration and vulnerability judgment, correctly identifying cases where the vulnerability persists despite exploit breakage.

The commit-based approaches show weaker performance. \vszzname achieves only 43.11\% in F1-score, with both low precision (58.34\%) and low recall (34.19\%). Its low precision indicates many false positives: \vszzname marks versions as affected based on commit ancestry, but not every version before a fix commit is actually vulnerable. Its low recall indicates that it misses many affected versions, likely because the introducing commit is not correctly identified or because multiple introducing commits exist. \llmname improves over \vszzname by using LLM-based commit analysis, raising F1-score to 69.95\% and precision to 84.63\%. This suggests that LLM-based commit screening reduces false introducing-commit identification. However, its recall remains low at 59.62\%, indicating that commit-based methods still inherently struggle to cover all affected versions comprehensively.

\subsubsection{Strength.}
Compared to \pocname, \appname demonstrates superior adaptability in handling cases where exploit execution fails for non-vulnerability reasons. When a exploit triggers an exception such as \texttt{NoSuchMethodError} or \texttt{ClassNotFoundException} on the target version, \pocname labels the version as \textit{not affected}, as the exploit did not successfully trigger. However, such failures are often caused by API-level changes that break the exploit harness rather than actual vulnerability fixes. \appname leverages trace-diff analysis to identify the specific code change that causes the exploit failure, and then determines whether the change removes the vulnerable path or merely breaks the exploit harness. This enables \appname to correctly label affected versions that \pocname misses.

Additionally, \appname leverages a structured evidence collection process that captures both fix evidence and exploit breakage evidence from version diffs. This design enables it to accommodate a broader range of code modifications related to vulnerability presence. This advantage becomes more prominent in complex vulnerability scenarios, such as deserialization and path traversal, where the fix involves subtle semantic changes that are difficult to detect from commit messages alone.

\subsubsection{Limitations.}
While \appname demonstrates strong performance in detecting vulnerability presence across library versions, it has several limitations. First, its recall on certain CWE types remains relatively low, particularly CWE-79 (64.24\%) and CWE-94 (68.18\%). These vulnerability types involve dynamic code generation and string manipulation, where the execution trace may not directly reveal the vulnerability trigger path. When the trace divergence does not align with the actual fix location, \appname may fail to collect sufficient evidence for a correct judgment.

Second, \appname encounters difficulties when the version diff contains a large number of changes that are unrelated to the vulnerability. In such cases, even with trace-diff guidance, the LLM may select evidence that appears relevant but does not accurately explain the trace divergence. This can lead to false positives when the selected evidence is incorrectly classified as exploit breakage rather than neutral.

\subsection{Ablation Study}

Our ablation study aims to achieve two goals: (1) to demonstrate that each component in our design contributes to the overall detection performance, and (2) to analyze how each component affects precision and recall differently. We construct four ablated variants of \appname:
\begin{itemize}[leftmargin=*]
    \item \textbf{\appname-tracediff (trace-free):} We remove the trace-diff summary and the version-diff overview from the initial prompt. The LLM only sees the failure excerpt from the target version. It must search for relevant diffs on its own. This ablation measures how much the trace-guided context helps.
    \item \textbf{\appname-diffsearch (random evidence):} We replace the LLM-selected evidence with randomly sampled diffs from the same version-diff space. This measures how much precise diff retrieval contributes.
    \item \textbf{\appname-aggregation (no rule):} We remove the rule-based label compression and use only the raw LLM judgment with simple threshold mapping. This measures how much the rule module contributes.
    \item \textbf{\appname-backtrack (no backfill):} We disable the version chain backfill module. Only the seed labels from the first two modules are used for evaluation. No forward or backward extension is performed along the version chain. This ablation measures how much the full-coverage backfill contributes.
\end{itemize}

All ablation variants use the same dataset, ground truth, and evaluation metrics as the full \appname pipeline. This ensures a fair comparison.

\begin{table}[htbp]
  \centering
  \caption{Ablation Study on \appname}
  \label{tab:ablation-appname}
  \resizebox{\columnwidth}{!}{
  \begin{tabular}{lcccccc}
    \toprule
    Tool & TP & FP & FN & Pre.(\%) & Rec.(\%) & F1(\%)\\
    \midrule
    \appname                & 11,741 & 335 & 1,368 & 97.23 & 89.56 & 93.24 \\
    \appname-tracediff      & 11,087 & 373 & 2,022 & 96.75 & 84.58 & 90.25 \\
    \appname-diffsearch     & 11,208 & 719 & 1,901 & 93.97 & 85.50 & 89.54 \\
    \appname-aggregation    & 11,433 & 320 & 1,676 & 97.28 & 87.21 & 91.97 \\
    \appname-backtrack      & 11,066 & 104 & 2,043 & 99.07 & 84.42 & 91.16 \\
    \bottomrule
  \end{tabular}
  }
\end{table}

\begin{table*}[t]
\centering
\caption{Results of Our Effectiveness Evaluation w.r.t CWE Types (\#V. denotes the number of vulnerabilities of a CWE type)}
\label{tab:cwe-results}
\resizebox{\linewidth}{!}{
\begin{tabular}{ll*{11}{c}}
\toprule
Tool & Metric &
\multicolumn{1}{c}{\begin{tabular}{c}CWE-502\\\#V.=8,082\end{tabular}} &
\multicolumn{1}{c}{\begin{tabular}{c}CWE-22\\\#V.=2,300\end{tabular}} &
\multicolumn{1}{c}{\begin{tabular}{c}CWE-611\\\#V.=1,687\end{tabular}} &
\multicolumn{1}{c}{\begin{tabular}{c}CWE-787\\\#V.=1,465\end{tabular}} &
\multicolumn{1}{c}{\begin{tabular}{c}CWE-770\\\#V.=1,094\end{tabular}} &
\multicolumn{1}{c}{\begin{tabular}{c}CWE-79\\\#V.=1,013\end{tabular}} &
\multicolumn{1}{c}{\begin{tabular}{c}CWE-20\\\#V.=773\end{tabular}} &
\multicolumn{1}{c}{\begin{tabular}{c}CWE-400\\\#V.=526\end{tabular}} &
\multicolumn{1}{c}{\begin{tabular}{c}CWE-94\\\#V.=499\end{tabular}} &
\multicolumn{1}{c}{\begin{tabular}{c}CWE-835\\\#V.=488\end{tabular}} &
\multicolumn{1}{c}{\begin{tabular}{c}OTHERS\\\#V.=8,783\end{tabular}} \\
\midrule
      & Pre.(\%) &
65.67 & 35.54 & 96.34 & 73.82 & 93.95 & 34.29 & 47.54 & 41.67 & 7.69 & 64.62 & 65.74 \\
\vszzname & Rec.(\%) &
11.33 & 61.92 & 17.73 & 24.12 & 38.32 & 55.63 & 20.71 & 37.69 & 11.93 & 87.26 & 58.58 \\
      & F1(\%) &
19.33 & 45.16 & 29.95 & 36.36 & 54.44 & 42.42 & 28.86 & 39.58 & 9.35 & 74.25 & 61.95 \\
\midrule
      & Pre.(\%) &
92.90 & 70.54 & 96.93 & 97.73 & 79.32 & 96.74 & 75.35 & 70.66 & 100.00 & 76.37 & 75.91 \\
\llmname & Rec.(\%) &
57.84 & 76.56 & 85.19 & 48.87 & 42.27 & 58.94 & 57.86 & 59.30 & 72.16 & 88.54 & 57.53 \\
      & F1(\%) &
71.30 & 73.42 & 90.68 & 65.15 & 55.15 & 73.25 & 65.45 & 64.48 & 83.83 & 82.01 & 65.45 \\
\midrule
      & Pre.(\%) &
99.52 & 100.00 & 100.00 & 98.16 & 100.00 & 89.39 & 100.00 & 98.41 & 100.00 & 100.00 & 99.15 \\
\pocname & Rec.(\%) &
90.32 & 84.55 & 86.08 & 93.51 & 73.68 & 52.98 & 80.71 & 93.47 & 57.95 & 87.90 & 78.80 \\
      & F1(\%) &
94.70 & 91.63 & 92.52 & 95.78 & 84.85 & 66.53 & 89.33 & 95.88 & 73.38 & 93.56 & 87.81 \\
\midrule
      & Pre.(\%) &
99.18 & 93.50 & 99.19 & 98.16 & 100.00 & 90.65 & 94.96 & 92.54 & 100.00 & 100.00 & 94.85 \\
\appname & Rec.(\%) &
92.59 & 87.67 & 95.85 & 93.51 & 75.66 & 64.24 & 80.71 & 93.47 & 68.18 & 93.63 & 88.71 \\
      & F1(\%) &
95.77 & 90.49 & 97.49 & 95.78 & 86.14 & 75.19 & 87.26 & 93.00 & 81.08 & 96.71 & 91.68 \\
\bottomrule
\end{tabular}}
\end{table*}

We observe from Table~\ref{tab:ablation-appname} that \appname achieves the highest F1-score, outperforming the best ablated variant by 1.27\%. This confirms that each component in our design contributes to the overall performance. Among the variants, removing the dynamic diff search causes the largest F1 drop of 3.70\%, followed by trace-diff guidance at 2.99\%, version chain backfill at 2.08\%, and rule-based aggregation at 1.27\%. Although the success rates degrade when removing any single module, they all remain notably higher than the commit-based baselines (\vszzname: 43.11\%, \llmname: 69.95\%), indicating that the core pipeline of \appname brings substantial benefits even without any single component.

\appname-tracediff removes the trace-diff summary and the version-diff overview from the initial prompt. Without trace-diff guidance, the LLM must start its search from the failure excerpt alone, making it harder to locate the correct diff hunks that explain the trace divergence. This results in the largest recall drop of 4.98\% among the first three variants, with 654 additional false negatives. The precision only drops slightly by 0.48\%, suggesting that the LLM can still make accurate judgments when it finds the right evidence, but it frequently fails to find it without the trace-diff context.

Beyond quantifying component importance, this variant also serves as a control against potential data leakage. Since the LLM was pre-trained on large-scale public data including CVE databases and GitHub repositories, it may have encountered some of the evaluated CVEs during pre-training. However, \appname-tracediff retains the CVE identifier in its prompt while removing trace-diff evidence; a model relying on memorized CVE-version associations would show minimal degradation under this condition. The substantial F1 and recall drops indicate that \appname genuinely depends on behavioral evidence rather than memorized CVE knowledge. The \appname-diffsearch variant reinforces this conclusion: replacing LLM-selected evidence with random diffs causes the largest F1 drop of 3.70\%, confirming that precise evidence retrieval matters beyond what the model could infer from the CVE identifier alone.


\appname-aggregation removes the rule-based label compression. The recall drops by 2.35\% while precision increases only slightly by 0.05\%. This is because the rule module helps recover borderline cases that the LLM alone classifies with low confidence, by considering additional signals such as trace quality, scenario type, and failure subtype that the raw LLM score alone does not capture.

\appname-backtrack disables the version chain backfill module. The recall drops by 5.14\%, the largest recall drop among all variants, while precision rises to 99.07\%. This trade-off is expected: the backfill module extends seed labels to uncovered versions along the version chain, increasing coverage at the cost of a small number of over-extensions. The 231 additional false positives are modest compared to the 675 recovered true positives. The high precision of the seed labels before backfill (99.07\%) also validates the quality of the first two modules' output.

\subsection{CWE-Level Analysis}

Table~\ref{tab:cwe-results} shows the per-CWE results for the top 10 CWE categories and an \textit{others} group.
\appname outperforms or matches all baselines in F1-score on 7 out of 11 CWE groups. The largest improvements over \pocname appear in CWE types where exploit execution frequently fails for non-vulnerability reasons. For CWE-94 (Code Injection), \appname achieves F1 of 81.08\% compared to 73.38\% for \pocname, a gain of 7.70\%. This is driven by a recall improvement from 57.95\% to 68.18\%. Code injection vulnerabilities often involve complex trigger paths that break across versions due to API changes rather than actual fixes. The trace-diff analysis of \appname can identify such cases and correctly label them as \textit{affected}. For CWE-79 (XSS), \appname improves F1 from 66.53\% to 75.19\%, with recall rising from 52.98\% to 64.24\%. XSS exploits are sensitive to minor changes in output encoding or input validation, which often break the exploit harness without fixing the underlying vulnerability.

The largest improvement over commit-based methods appears in CWE-502 (Deserialization), the largest CWE group with 8,082 vulnerabilities. \appname achieves F1 of 95.77\%, compared to 71.30\% for \llmname and 19.33\% for \vszzname. Deserialization vulnerabilities often span multiple library versions with subtle changes in serialization logic. Commit-based methods struggle because the relevant fix commits are not easily identifiable from commit messages alone. \vszzname achieves particularly low precision (65.67\%) and recall (11.33\%) on CWE-502, indicating that both fix-commit identification and version coverage are problematic for this vulnerability type.


The CWE-level results also reveal a clear divide between commit-based methods and execution-based methods. \vszzname performs poorly on most CWE types, with F1 below 50.00\% on 8 out of 11 groups. Its lowest F1 is 9.35\% on CWE-94, where commit analysis fails to identify code injection fixes. \llmname improves over \vszzname but still lags behind execution-based methods on most CWE types. Interestingly, \llmname achieves competitive F1 on CWE-94 (83.83\%) compared to \pocname (73.38\%), suggesting that LLM-based commit analysis is particularly effective for code injection vulnerabilities where fix commits have clear semantic signals. However, \llmname struggles with CWE-770 (F1 = 55.15\%) and CWE-787 (F1 = 65.15\%), where the relationship between commits and vulnerability presence is less direct.

\section{Discussion}
\label{sec:discussion}


This section reflects on what we have learned from the study.
We first highlight key success cases, including comparisons with exploit-only baselines and a patch-based approach.
We then summarize typical failure cases and their root causes.
After that, we discuss the cost of using large language models.
Finally, we outline the main threats to the validity of our findings.

\subsection{Success Cases}

\appname achieves an F1-score of 93.24\% across 25,943 versions, outperforming commit-based and exploit-only baselines. The 2.34\% F1 gain over the exploit-only baseline translates to 620 additional affected versions identified—versions where the exploit fails to trigger but the vulnerability persists. While GitHub Advisory Database already covers 92.6\% of these, 46 versions across five CVEs remain unlisted in any advisory. For instance, CVE-2022-25845 (Fastjson) affects versions from 1.2.10 to 1.2.24, yet GitHub Advisory omits 34 early versions in this range. The exploit baseline misses them because the exploit fails to build on these versions due to API changes; \appname correctly recovers them through trace-diff analysis. On the 82 CVEs common with VISION~\cite{Wu2024Vision} (12,041 versions), \appname attains 93.27\% F1 versus VISION's 85.75\%, as \appname anchors analysis on exploit behavior rather than patch structure, avoiding over-approximation when introducing and fixing commits are structurally distant.

\subsection{Qualitative Failure Analysis}
\label{subsec:discussion-failure}

Although the overall performance is strong, we also find clear limits.
We group the most common failure patterns into three types.
These patterns help us understand where \appname still needs improvement.

\textbf{Fragile test and environment.}
Failures in the trace construction module (Module~1) arise when the exploit cannot compile or run on the target version, leaving \appname with little to no execution trace. For example, in CVE-2013-7285 (XStream), the exploit fails to build on version~1.0 due to API incompatibility, so \appname has no behavioral signal and remains unlabeled across many early versions. These cases highlight that the current pipeline does not repair build or dependency issues to recover trace signals.

\textbf{Misinterpretation of evidence.}
Failures in the evidence typing and case-level judgment modules (Module~2) occur when the LLM correctly locates relevant code changes but draws the wrong conclusion. For instance, in CVE-2024-1597 (PostgreSQL JDBC), \appname identifies code differences and confidently judges the vulnerability as fixed, when in reality the affected logic persists under a different execution path. Such cases show that the model can still misattribute changes as a complete vulnerability removal.

\textbf{Gaps along version chains.}
Failures in the version chain backfill module (Module~3) arise from incomplete coverage. For example, CVE-2022-25845 (Fastjson) has 110 versions where \appname produces no label because the seed labels from difftrace never reached these boundary versions, and the backfill rules conservatively avoid propagating labels across execution-outcome boundaries. This trade-off between completeness and caution is inherent to the region-based propagation design.



\subsection{Threats to Validity}
\label{subsec:discussion-threats}

The main threats to validity come from the nature of version-level vulnerability analysis in \appname.

\textbf{External validity.}
Our study focuses on Java library vulnerabilities with runnable exploits, recoverable releases, and accessible version histories.
Although this setting is broad, it does not cover many real-world cases, such as closed-source software, deployed applications, or projects with missing history.
The reliance on publicly available exploits also limits the scope to vulnerabilities for which a working PoC exists.
Therefore, our findings may not directly generalize to all software ecosystems.

\textbf{Internal validity.}
Another threat lies in how we decide whether a version is affected.
This is not always directly visible: a version may still contain the bug even if one exploit no longer works, and the result of an exploit may change with configuration, dependencies, or runtime environment.
In addition, datasets usually provide only a small number of exploits and limited historical evidence.
The LLM judgments may also vary across different prompts or model versions.
As a result, some affected-version labels may still be uncertain.
This problem is common in version-level vulnerability studies and is not specific to our approach.





\section{Related Work}
\label{sec:related}

\textbf{Exploit-Based Analysis.} Exploit migration adapts existing exploits to different software versions for vulnerability assessment and reproduction. AEM~\cite{Jiang2023AEM} aligns execution points across Linux kernel versions to reproduce exploitation behaviors. VulScope~\cite{Dai2021Exploit} leverages directed fuzzing to migrate exploits between versions. SyzBridge~\cite{zou2024SyzBridge} addresses environmental differences between upstream and downstream kernels. Evocatio~\cite{Jiang2022Expand} automatically generates exploits to expose unknown bug capabilities. These approaches rely on fuzzing and explicit execution trace mapping, which is time-consuming and challenging. Dai et al. utilized exploit migration with directed fuzzing to identify affected versions, but the vulnerability may lack a exploit~\cite{Mu2018Reproduce,Dong2019Version}. Moreover, existing exploit-based approaches treat execution failure as evidence that the vulnerability is absent, without distinguishing genuine fixes from exploit breakages caused by environmental changes. Exploit migration also shares similarities with test migration~\cite{Alshahwan2024meta,gazzola2018automatic,Saboor2025repair,rahman2025utfix,Rahman2024flaky,Li2019repair} and API migration~\cite{dig2006apis,dagenais2011recommending}, which can partially address exploit failures caused by API changes but typically focus on function-level repair and overlook environmental factors. 

\textbf{Affected Version Identification.} Various approaches have been proposed to identify library versions affected by an OSS vulnerability~\cite{Dong2019Version,Bao2022Vszz,nguyen2013reliability,nguyen2016automatic,shi2022precise,He2024Logs,xiao2020mvp}. Report-based methods such as NER~\cite{Dong2019Version} extract version information from vulnerability reports, but are limited by report quality and completeness~\cite{Dong2019Version,nguyen2013reliability,Mu2018Reproduce}. SZZ-based approaches trace vulnerability-introducing changes through patch analysis: the original SZZ~\cite{sliwerski2005changes} and its improvement V-SZZ~\cite{Bao2022Vszz} backtrack deleted lines in patches to identify vulnerability-inducing versions, but fail when patches only contain added lines. Patch-based approaches~\cite{He2024Logs,shi2022precise} identify affected versions by measuring patch line presence in target versions or leveraging developer logs, but they ignore the context of modified lines and require manual verification. Vision~\cite{Wu2024Vision} captures patch context by encoding critical methods and statements into weighted inter-procedural dependency graph signatures. VFC analysis methods~\cite{li2016vulpecker,li2021sysevr,li2018vuldeepecker} extract vulnerability features from fixing commits, but rely heavily on predefined rules. VERCATION~\cite{cheng2025vercation} leverages LLMs and AST-based code clone detection to improve vulnerability characterization. Vulnerable code clone detection approaches such as VUDDY~\cite{kim2017vuddy}, MVP~\cite{xiao2020mvp}, MOVERY~\cite{woo2022movery}, V1SCAN~\cite{woo2023v1scan}, V0Finder~\cite{woo2021v0finder}, and VCCFinder~\cite{perl2015vccfinder} generate syntactic or semantic signatures to detect recurring vulnerabilities, but these signatures often contain irrelevant code. 



\section{Conclusion}
\label{sec:conclusion}

In this work, we presented \appname, a trace-driven diff analysis framework for automated exploit failure analysis across evolving library versions. It comprises three modules: trace construction builds version context from exploit executions, diff exploration uses a tool-augmented LLM loop to collect vulnerability evidence from diffs, and affected-version judgment applies version-chain backfill to label versions. On 224 CVEs and 25,943 versions from 128 Java libraries, \appname achieves an F1-score of 93.24\%, outperforming \vszzname and \llmname by 116.28\% and 33.30\% respectively, while keeping token costs low via short trace-guided prompts and bounded iterations. Ablation and CWE analyses show that trace-diff guidance, dynamic diff search, rule-based aggregation, and version-chain backfill all contribute to effectiveness, particularly when exploits fail for non-vulnerability reasons or commits do not clearly delimit affected versions. \appname still faces challenges on vulnerability types with weak or noisy trace signals and in tangled diffs. Future work may explore richer trace collection, stronger robustness to unrelated changes, and broader application beyond Java. Overall, \appname makes exploit-based vulnerability assessment more reliable across versions.

\section*{Acknowledgement}
This research is supported by the Fundamental Research Funds for the Central Universities (No. 226-2025-00171). We also thank the anonymous reviewers for their insightful suggestions.

\balance
\bibliographystyle{ACM-Reference-Format}
\bibliography{main}

@inproceedings{Dong2019Version,
author = {Dong, Ying and Guo, Wenbo and Chen, Yueqi and Xing, Xinyu and Zhang, Yuqing and Wang, Gang},
title = {Towards the detection of inconsistencies in public security vulnerability reports},
year = {2019},
isbn = {9781939133069},
publisher = {USENIX Association},
address = {USA},
booktitle = {Proceedings of the 28th USENIX Conference on Security Symposium},
pages = {869–885},
numpages = {17},
location = {Santa Clara, CA, USA},
series = {SEC'19}
}

@inproceedings{zirui2024exploiting,
author = {Chen, Zirui and Hu, Xing and Xia, Xin and Gao, Yi and Xu, Tongtong and Lo, David and Yang, Xiaohu},
title = {Exploiting Library Vulnerability via Migration Based Automating Test Generation},
year = {2024},
isbn = {9798400702174},
publisher = {Association for Computing Machinery},
address = {New York, NY, USA},
url = {https://doi.org/10.1145/3597503.3639583},
doi = {10.1145/3597503.3639583},
booktitle = {Proceedings of the IEEE/ACM 46th International Conference on Software Engineering},
articleno = {228},
numpages = {12},
location = {Lisbon, Portugal},
series = {ICSE '24}
}

@inproceedings{Mu2018Reproduce,
author = {Mu, Dongliang and Cuevas, Alejandro and Yang, Limin and Hu, Hang and Xing, Xinyu and Mao, Bing and Wang, Gang},
title = {Understanding the reproducibility of crowd-reported security vulnerabilities},
year = {2018},
isbn = {9781931971461},
publisher = {USENIX Association},
address = {USA},
booktitle = {Proceedings of the 27th USENIX Conference on Security Symposium},
pages = {919–936},
numpages = {18},
location = {Baltimore, MD, USA},
series = {SEC'18}
}

@INPROCEEDINGS {Li2019repair,
author = { Li, Xiangyu and d'Amorim, Marcelo and Orso, Alessandro },
booktitle = { 2019 12th IEEE Conference on Software Testing, Validation and Verification (ICST) },
title = {{ Intent-Preserving Test Repair }},
year = {2019},
volume = {},
ISSN = {2159-4848},
pages = {217-227},
doi = {10.1109/ICST.2019.00030},
url = {https://doi.ieeecomputersociety.org/10.1109/ICST.2019.00030},
publisher = {IEEE Computer Society},
address = {Los Alamitos, CA, USA},
month =apr}

@inproceedings{Dai2021Exploit,
author = {Dai, Jiarun and Zhang, Yuan and Xu, Hailong and Lyu, Haiming and Wu, Zicheng and Xing, Xinyu and Yang, Min},
title = {Facilitating Vulnerability Assessment through PoC Migration},
year = {2021},
isbn = {9781450384544},
publisher = {Association for Computing Machinery},
address = {New York, NY, USA},
url = {https://doi.org/10.1145/3460120.3484594},
doi = {10.1145/3460120.3484594},
booktitle = {Proceedings of the 2021 ACM SIGSAC Conference on Computer and Communications Security},
pages = {3300–3317},
numpages = {18},
location = {Virtual Event, Republic of Korea},
series = {CCS '21}
}

@inproceedings{Jiang2022Expand,
author = {Jiang, Zhiyuan and Gan, Shuitao and Herrera, Adrian and Toffalini, Flavio and Romerio, Lucio and Tang, Chaojing and Egele, Manuel and Zhang, Chao and Payer, Mathias},
title = {Evocatio: Conjuring Bug Capabilities from a Single PoC},
year = {2022},
isbn = {9781450394505},
publisher = {Association for Computing Machinery},
address = {New York, NY, USA},
url = {https://doi.org/10.1145/3548606.3560575},
doi = {10.1145/3548606.3560575},
booktitle = {Proceedings of the 2022 ACM SIGSAC Conference on Computer and Communications Security},
pages = {1599–1613},
numpages = {15},
location = {Los Angeles, CA, USA},
series = {CCS '22}
}

@inproceedings{Alshahwan2024meta,
author = {Alshahwan, Nadia and Chheda, Jubin and Finogenova, Anastasia and Gokkaya, Beliz and Harman, Mark and Harper, Inna and Marginean, Alexandru and Sengupta, Shubho and Wang, Eddy},
title = {Automated Unit Test Improvement using Large Language Models at Meta},
year = {2024},
isbn = {9798400706585},
publisher = {Association for Computing Machinery},
address = {New York, NY, USA},
url = {https://doi.org/10.1145/3663529.3663839},
doi = {10.1145/3663529.3663839},
booktitle = {Companion Proceedings of the 32nd ACM International Conference on the Foundations of Software Engineering},
pages = {185–196},
numpages = {12},
location = {Porto de Galinhas, Brazil},
series = {FSE 2024}
}

@inproceedings{Rahman2024flaky,
author = {Rahman, Shanto and Shi, August},
title = {FlakeSync: Automatically Repairing Async Flaky Tests},
year = {2024},
isbn = {9798400702174},
publisher = {Association for Computing Machinery},
address = {New York, NY, USA},
url = {https://doi.org/10.1145/3597503.3639115},
doi = {10.1145/3597503.3639115},
booktitle = {Proceedings of the IEEE/ACM 46th International Conference on Software Engineering},
articleno = {136},
numpages = {12},
location = {Lisbon, Portugal},
series = {ICSE '24}
}

@article{Li2023downstream,
author = {Li, Siyuan and Wang, Yongpan and Dong, Chaopeng and Yang, Shouguo and Li, Hong and Sun, Hao and Lang, Zhe and Chen, Zuxin and Wang, Weijie and Zhu, Hongsong and Sun, Limin},
title = {LibAM: An Area Matching Framework for Detecting Third-Party Libraries in Binaries},
year = {2023},
issue_date = {February 2024},
publisher = {Association for Computing Machinery},
address = {New York, NY, USA},
volume = {33},
number = {2},
issn = {1049-331X},
url = {https://doi.org/10.1145/3625294},
doi = {10.1145/3625294},
journal = {ACM Trans. Softw. Eng. Methodol.},
month = dec,
articleno = {52},
numpages = {35}
}

@INPROCEEDINGS{Jiang2023AEM,
  author={Jiang, Zheyue and Zhang, Yuan and Xu, Jun and Sun, Xinqian and Liu, Zhuang and Yang, Min},
  booktitle={2023 IEEE Symposium on Security and Privacy (SP)}, 
  title={AEM: Facilitating Cross-Version Exploitability Assessment of Linux Kernel Vulnerabilities}, 
  year={2023},
  volume={},
  number={},
  pages={2122-2137},
  doi={10.1109/SP46215.2023.10179286}}

@article{zou2024SyzBridge,
place = {Country unknown/Code not available}, title = {SyzBridge: Bridging the Gap in Exploitability Assessment of Linux Kernel Bugs in the Linux Ecosystem}, 
year = {2024},
url = {https://par.nsf.gov/biblio/10493939}, DOI = {10.14722/ndss.2024.24926}, abstractNote = {}, journal = {NDSS}, publisher = {Internet Society}, author = {Zou, Xiaochen and Hao, Yu and Zhang, Zheng and Pu, Juefei and Chen, Weiteng and Qian, Zhiyun}, }

@inproceedings{Zhou2024Magneto,
author = {Zhou, Zhuotong and Yang, Yongzhuo and Wu, Susheng and Huang, Yiheng and Chen, Bihuan and Peng, Xin},
title = {Magneto: A Step-Wise Approach to Exploit Vulnerabilities in Dependent Libraries via LLM-Empowered Directed Fuzzing},
year = {2024},
isbn = {9798400712487},
publisher = {Association for Computing Machinery},
address = {New York, NY, USA},
url = {https://doi.org/10.1145/3691620.3695531},
doi = {10.1145/3691620.3695531},
booktitle = {Proceedings of the 39th IEEE/ACM International Conference on Automated Software Engineering},
pages = {1633–1644},
numpages = {12},
location = {Sacramento, CA, USA},
series = {ASE '24}
}

@inproceedings{Bao2022Vszz,
author = {Bao, Lingfeng and Xia, Xin and Hassan, Ahmed E. and Yang, Xiaohu},
title = {V-SZZ: automatic identification of version ranges affected by CVE vulnerabilities},
year = {2022},
isbn = {9781450392211},
publisher = {Association for Computing Machinery},
address = {New York, NY, USA},
url = {https://doi.org/10.1145/3510003.3510113},
doi = {10.1145/3510003.3510113},
booktitle = {Proceedings of the 44th International Conference on Software Engineering},
pages = {2352–2364},
numpages = {13},
location = {Pittsburgh, Pennsylvania},
series = {ICSE '22}
}

@inproceedings{Wu2024Vision,
author = {Wu, Susheng and Wang, Ruisi and Huang, Kaifeng and Cao, Yiheng and Song, Wenyan and Zhou, Zhuotong and Huang, Yiheng and Chen, Bihuan and Peng, Xin},
title = {Vision: Identifying Affected Library Versions for Open Source Software Vulnerabilities},
year = {2024},
isbn = {9798400712487},
publisher = {Association for Computing Machinery},
address = {New York, NY, USA},
url = {https://doi.org/10.1145/3691620.3695516},
doi = {10.1145/3691620.3695516},
booktitle = {Proceedings of the 39th IEEE/ACM International Conference on Automated Software Engineering},
pages = {1447–1459},
numpages = {13},
location = {Sacramento, CA, USA},
series = {ASE '24}
}

@ARTICLE{He2024Logs,
  author={He, Yongzhong and Wang, Yiming and Zhu, Sencun and Wang, Wei and Zhang, Yunjia and Li, Qiang and Yu, Aimin},
  journal={IEEE Transactions on Dependable and Secure Computing}, 
  title={Automatically Identifying CVE Affected Versions With Patches and Developer Logs}, 
  year={2024},
  volume={21},
  number={2},
  pages={905-919},
  doi={10.1109/TDSC.2023.3264567}}

@article{shuhan2025vul,
author = {Liu, Shuhan and Zhou, Jiayuan and Hu, Xing and Cogo, Filipe Roseiro and Xia, Xin and Yang, Xiaohu},
title = {An Empirical Study on Vulnerability Disclosure Management of Open Source Software Systems},
year = {2025},
issue_date = {September 2025},
publisher = {Association for Computing Machinery},
address = {New York, NY, USA},
volume = {34},
number = {7},
issn = {1049-331X},
url = {https://doi.org/10.1145/3716822},
doi = {10.1145/3716822},
journal = {ACM Trans. Softw. Eng. Methodol.},
month = aug,
articleno = {214},
numpages = {31}
}

@misc{gao2026depradaragenticcoordinationcontext,
      title={DepRadar: Agentic Coordination for Context Aware Defect Impact Analysis in Deep Learning Libraries}, 
      author={Yi Gao and Xing Hu and Tongtong Xu and Jiali Zhao and Xiaohu Yang and Xin Xia},
      year={2026},
      eprint={2601.09440},
      archivePrefix={arXiv},
      primaryClass={cs.SE},
      url={https://arxiv.org/abs/2601.09440}, 
}

@article{Gao2025LLM,
  title={Automated unit test refactoring},
  author={Gao, Yi and Hu, Xing and Yang, Xiaohu and Xia, Xin},
  journal={Proceedings of the ACM on Software Engineering},
  volume={2},
  number={FSE},
  pages={713--733},
  year={2025},
  publisher={ACM New York, NY, USA}
}

@misc{chen2025diffploitfacilitatingcrossversionexploit,
      title={Diffploit: Facilitating Cross-Version Exploit Migration for Open Source Library Vulnerabilities}, 
      author={Zirui Chen and Zhipeng Xue and Jiayuan Zhou and Xing Hu and Xin Xia and Xiaohu Yang},
      year={2025},
      eprint={2511.12950},
      archivePrefix={arXiv},
      primaryClass={cs.SE},
      url={https://arxiv.org/abs/2511.12950}, 
}

@misc{li2022poisonattackdefensedeep,
      title={Poison Attack and Defense on Deep Source Code Processing Models}, 
      author={Jia Li and Zhuo Li and Huangzhao Zhang and Ge Li and Zhi Jin and Xing Hu and Xin Xia},
      year={2022},
      eprint={2210.17029},
      archivePrefix={arXiv},
      primaryClass={cs.SE},
      url={https://arxiv.org/abs/2210.17029}, 
}

@misc{chen2025generatingmitigationsdownstreamprojects,
      title={Generating Mitigations for Downstream Projects to Neutralize Upstream Library Vulnerability}, 
      author={Zirui Chen and Xing Hu and Puhua Sun and Xin Xia and Xiaohu Yang},
      year={2025},
      eprint={2503.24273},
      archivePrefix={arXiv},
      primaryClass={cs.SE},
      url={https://arxiv.org/abs/2503.24273}, 
}

@misc{yu2025selfadmittedcodegeneratedlarge,
      title={Where Is Self-admitted Code Generated by Large Language Models on GitHub?}, 
      author={Xiao Yu and Lei Liu and Xing Hu and Jin Liu and Xin Xia},
      year={2025},
      eprint={2406.19544},
      archivePrefix={arXiv},
      primaryClass={cs.SE},
      url={https://arxiv.org/abs/2406.19544}, 
}

@article{Xue2025LLM,
author = {Xue, Zhipeng and Zhang, Xiaoting and Gao, Zhipeng and Hu, Xing and Gao, Shan and Xia, Xin and Li, Shanping},
title = {Clean Code, Better Models: Enhancing LLM Performance with Smell-Cleaned Dataset},
year = {2026},
publisher = {Association for Computing Machinery},
address = {New York, NY, USA},
issn = {1049-331X},
url = {https://doi.org/10.1145/3793252},
doi = {10.1145/3793252},
note = {Just Accepted},
journal = {ACM Trans. Softw. Eng. Methodol.},
month = feb
}

@inproceedings{Xue2024LLM,
author = {Xue, Zhipeng and Gao, Zhipeng and Wang, Shaohua and Hu, Xing and Xia, Xin and Li, Shanping},
title = {SelfPiCo: Self-Guided Partial Code Execution with LLMs},
year = {2024},
isbn = {9798400706127},
publisher = {Association for Computing Machinery},
address = {New York, NY, USA},
url = {https://doi.org/10.1145/3650212.3680368},
doi = {10.1145/3650212.3680368},
booktitle = {Proceedings of the 33rd ACM SIGSOFT International Symposium on Software Testing and Analysis},
pages = {1389–1401},
numpages = {13},
location = {Vienna, Austria},
series = {ISSTA 2024}
}

@online{introduce51080,
    title = {Introducing Commit of CVE-2023-51080},
    url = {https://github.com/chinabugotech/hutool/commit/c45b3f},
    author = {Looly},
}

@misc{replication,
  author       = {Xinwei Mao},
  title        = {{ATTAIN}: Automated Exploit Failure Analysis through Trace-Driven Diff Analysis},
  year         = {2026},
  month        = apr,
  howpublished = {GitHub repository},
  url          = {https://github.com/Cirno-9-lab/ATTAIN_REPLICATION}
}

@inproceedings{jiayuan2023cole,
author = {Zhou, Jiayuan and Pacheco, Michael and Chen, Jinfu and Hu, Xing and Xia, Xin and Lo, David and Hassan, Ahmed E.},
title = {CoLeFunDa: Explainable Silent Vulnerability Fix Identification},
year = {2023},
isbn = {9781665457019},
publisher = {IEEE Press},
url = {https://doi.org/10.1109/ICSE48619.2023.00214},
doi = {10.1109/ICSE48619.2023.00214},
booktitle = {Proceedings of the 45th International Conference on Software Engineering},
pages = {2565–2577},
numpages = {13},
location = {Melbourne, Victoria, Australia},
series = {ICSE '23}
}

@inproceedings{jiayuan2022fix,
author = {Zhou, Jiayuan and Pacheco, Michael and Wan, Zhiyuan and Xia, Xin and Lo, David and Wang, Yuan and Hassan, Ahmed E.},
title = {Finding a needle in a haystack: automated mining of silent vulnerability fixes},
year = {2022},
isbn = {9781665403375},
publisher = {IEEE Press},
url = {https://doi.org/10.1109/ASE51524.2021.9678720},
doi = {10.1109/ASE51524.2021.9678720},
booktitle = {Proceedings of the 36th IEEE/ACM International Conference on Automated Software Engineering},
pages = {705–716},
numpages = {12},
location = {Melbourne, Australia},
series = {ASE '21}
}

@article{tang2025llm4szz,
  title={LLM4SZZ: Enhancing szz algorithm with context-enhanced assessment on large language models},
  author={Tang, Lingxiao and Liu, Jiakun and Liu, Zhongxin and Yang, Xiaohu and Bao, Lingfeng},
  journal={Proceedings of the ACM on Software Engineering},
  volume={2},
  number={ISSTA},
  pages={343--365},
  year={2025},
  publisher={ACM New York, NY, USA}
}

@inproceedings{gazzola2018automatic,
  title={Automatic software repair: A survey},
  author={Gazzola, Luca and Micucci, Daniela and Mariani, Leonardo},
  booktitle={Proceedings of the 40th International Conference on Software Engineering},
  pages={1219--1219},
  year={2018}
}

@article{Saboor2025repair,
author = {Saboor Yaraghi, Ahmadreza and Holden, Darren and Kahani, Nafiseh and Briand, Lionel},
title = {Automated Test Case Repair Using Language Models},
year = {2025},
issue_date = {April 2025},
publisher = {IEEE Press},
volume = {51},
number = {4},
issn = {0098-5589},
url = {https://doi.org/10.1109/TSE.2025.3541166},
doi = {10.1109/TSE.2025.3541166},
journal = {IEEE Trans. Softw. Eng.},
month = feb,
pages = {1104–1133},
numpages = {30}
}

@article{Kula1,
  title={Do developers update their library dependencies? An empirical study on the impact of security advisories on library migration},
  author={Kula, Raula Gaikovina and German, Daniel M and Ouni, Ali and Ishio, Takashi and Inoue, Katsuro},
  journal={Empirical Software Engineering},
  volume={23},
  pages={384--417},
  year={2018},
  publisher={Springer}
}

@inproceedings{Markus1,
author = {Markus Zimmermann and Cristian-Alexandru Staicu and Cam Tenny and Michael Pradel},
title = {Small World with High Risks: A Study of Security Threats in the npm Ecosystem},
booktitle = {28th USENIX Security Symposium (USENIX Security 19)},
year = {2019},
isbn = {978-1-939133-06-9},
address = {Santa Clara, CA},
pages = {995--1010},
publisher = {USENIX Association},
month = aug
}

@online{Synopsys1,
    title = {{OPEN} {SOURCE} {SECURITY} {AND} {RISK} {ANALYSIS} {REPORT} 2023},
    url = {https://www.synopsys.com/software-integrity/resources/analyst-reports/open-source-security-risk-analysis.html},
    author = {Synopsys},
}

@article{Bavota1,
  title={How the apache community upgrades dependencies: an evolutionary study},
  author={Bavota, Gabriele and Canfora, Gerardo and Di Penta, Massimiliano and Oliveto, Rocco and Panichella, Sebastiano},
  journal={Empirical Software Engineering},
  volume={20},
  pages={1275--1317},
  year={2015},
  publisher={Springer}
}

@article{kula2018developers,
  title={Do developers update their library dependencies? An empirical study on the impact of security advisories on library migration},
  author={Kula, Raula Gaikovina and German, Daniel M and Ouni, Ali and Ishio, Takashi and Inoue, Katsuro},
  journal={Empirical Software Engineering},
  volume={23},
  pages={384--417},
  year={2018},
  publisher={Springer}
}

@article{He2023Dependent,
author = {He, Runzhi and He, Hao and Zhang, Yuxia and Zhou, Minghui},
title = {Automating Dependency Updates in Practice: An Exploratory Study on GitHub Dependabot},
year = {2023},
issue_date = {Aug. 2023},
publisher = {IEEE Press},
volume = {49},
number = {8},
issn = {0098-5589},
url = {https://doi.org/10.1109/TSE.2023.3278129},
doi = {10.1109/TSE.2023.3278129},
journal = {IEEE Trans. Softw. Eng.},
month = aug,
pages = {4004–4022},
numpages = {19}
}

@inproceedings{Croft2023Report,
author = {Croft, Roland and Babar, M. Ali and Kholoosi, M. Mehdi},
title = {Data Quality for Software Vulnerability Datasets},
year = {2023},
isbn = {9781665457019},
publisher = {IEEE Press},
url = {https://doi.org/10.1109/ICSE48619.2023.00022},
doi = {10.1109/ICSE48619.2023.00022},
booktitle = {Proceedings of the 45th International Conference on Software Engineering},
pages = {121–133},
numpages = {13},
location = {Melbourne, Victoria, Australia},
series = {ICSE '23}
}

@ARTICLE{Jo2021Report,
  author={Jo, Hyeonseong and Kim, Jinwoo and Porras, Phillip and Yegneswaran, Vinod and Shin, Seungwon},
  journal={IEEE Transactions on Information Forensics and Security}, 
  title={GapFinder: Finding Inconsistency of Security Information From Unstructured Text}, 
  year={2021},
  volume={16},
  number={},
  pages={86-99},
  doi={10.1109/TIFS.2020.3003570}}

@misc{grotov2024untanglingknotsleveragingllm,
      title={Untangling Knots: Leveraging LLM for Error Resolution in Computational Notebooks}, 
      author={Konstantin Grotov and Sergey Titov and Yaroslav Zharov and Timofey Bryksin},
      year={2024},
      eprint={2405.01559},
      archivePrefix={arXiv},
      primaryClass={cs.SE},
      url={https://arxiv.org/abs/2405.01559}, 
}

@misc{zou2024docbenchbenchmarkevaluatingllmbased,
      title={DOCBENCH: A Benchmark for Evaluating LLM-based Document Reading Systems}, 
      author={Anni Zou and Wenhao Yu and Hongming Zhang and Kaixin Ma and Deng Cai and Zhuosheng Zhang and Hai Zhao and Dong Yu},
      year={2024},
      eprint={2407.10701},
      archivePrefix={arXiv},
      primaryClass={cs.CL},
      url={https://arxiv.org/abs/2407.10701}, 
}

@misc{tyen2024llmsreasoningerrorscorrect,
      title={LLMs cannot find reasoning errors, but can correct them given the error location}, 
      author={Gladys Tyen and Hassan Mansoor and Victor Cărbune and Peter Chen and Tony Mak},
      year={2024},
      eprint={2311.08516},
      archivePrefix={arXiv},
      primaryClass={cs.AI},
      url={https://arxiv.org/abs/2311.08516}, 
}

@INPROCEEDINGS{Pan2023Type,
  author={Pan, Shengyi and Bao, Lingfeng and Xia, Xin and Lo, David and Li, Shanping},
  booktitle={2023 IEEE/ACM 45th International Conference on Software Engineering (ICSE)}, 
  title={Fine-grained Commit-level Vulnerability Type Prediction by CWE Tree Structure}, 
  year={2023},
  volume={},
  number={},
  pages={957-969},
  doi={10.1109/ICSE48619.2023.00088}}

@inproceedings{Pan2022Reports,
author = {Pan, Shengyi and Zhou, Jiayuan and Cogo, Filipe Roseiro and Xia, Xin and Bao, Lingfeng and Hu, Xing and Li, Shanping and Hassan, Ahmed E.},
title = {Automated unearthing of dangerous issue reports},
year = {2022},
isbn = {9781450394130},
publisher = {Association for Computing Machinery},
address = {New York, NY, USA},
url = {https://doi.org/10.1145/3540250.3549156},
doi = {10.1145/3540250.3549156},
booktitle = {Proceedings of the 30th ACM Joint European Software Engineering Conference and Symposium on the Foundations of Software Engineering},
pages = {834–846},
numpages = {13},
location = {Singapore, Singapore},
series = {ESEC/FSE 2022}
}

@inproceedings{Pan2024Assessment,
author = {Pan, Shengyi and Bao, Lingfeng and Zhou, Jiayuan and Hu, Xing and Xia, Xin and Li, Shanping},
title = {Towards More Practical Automation of Vulnerability Assessment},
year = {2024},
isbn = {9798400702174},
publisher = {Association for Computing Machinery},
address = {New York, NY, USA},
url = {https://doi.org/10.1145/3597503.3639110},
doi = {10.1145/3597503.3639110},
booktitle = {Proceedings of the IEEE/ACM 46th International Conference on Software Engineering},
articleno = {148},
numpages = {13},
location = {Lisbon, Portugal},
series = {ICSE '24}
}

@inproceedings{Zhan2024PS3,
author = {Zhan, Qi and Hu, Xing and Li, Zhiyang and Xia, Xin and Lo, David and Li, Shanping},
title = {PS3: Precise Patch Presence Test based on Semantic Symbolic Signature},
year = {2024},
isbn = {9798400702174},
publisher = {Association for Computing Machinery},
address = {New York, NY, USA},
url = {https://doi.org/10.1145/3597503.3639134},
doi = {10.1145/3597503.3639134},
booktitle = {Proceedings of the IEEE/ACM 46th International Conference on Software Engineering},
articleno = {167},
numpages = {12},
location = {Lisbon, Portugal},
series = {ICSE '24}
}

@inproceedings{Zhan2025React,
author = {Zhan, Qi and Hu, Xing and Xia, Xin and Li, Shanping},
title = {REACT: IR-Level Patch Presence Test for Binary},
year = {2024},
isbn = {9798400712487},
publisher = {Association for Computing Machinery},
address = {New York, NY, USA},
url = {https://doi.org/10.1145/3691620.3695012},
doi = {10.1145/3691620.3695012},
booktitle = {Proceedings of the 39th IEEE/ACM International Conference on Automated Software Engineering},
pages = {381–392},
numpages = {12},
location = {Sacramento, CA, USA},
series = {ASE '24}
}

@article{Rahman2025UTFix,
author = {Rahman, Shanto and Kuhar, Sachit and Cirisci, Berk and Garg, Pranav and Wang, Shiqi and Ma, Xiaofei and Deoras, Anoop and Ray, Baishakhi},
title = {UTFix: Change Aware Unit Test Repairing using LLM},
year = {2025},
issue_date = {April 2025},
publisher = {Association for Computing Machinery},
address = {New York, NY, USA},
volume = {9},
number = {OOPSLA1},
url = {https://doi.org/10.1145/3720419},
doi = {10.1145/3720419},
journal = {Proc. ACM Program. Lang.},
month = apr,
articleno = {85},
numpages = {26}
}

@misc{chen2026largescaleempiricalstudygeneralizability,
      title={A Large-scale Empirical Study on the Generalizability of Disclosed Java Library Vulnerability Exploits}, 
      author={Zirui Chen and Qi Zhan and Jiayuan Zhou and Xing Hu and Xin Xia and Xiaohu Yang},
      year={2026},
      eprint={2603.25997},
      archivePrefix={arXiv},
      primaryClass={cs.SE},
      url={https://arxiv.org/abs/2603.25997}, 
}

@article{dig2006apis,
  title={How do APIs evolve? A story of refactoring},
  author={Dig, Danny and Johnson, Ralph},
  journal={Journal of software maintenance and evolution: Research and Practice},
  volume={18},
  number={2},
  pages={83--107},
  year={2006},
  publisher={Wiley Online Library}
}

@article{dagenais2011recommending,
  title={Recommending adaptive changes for framework evolution},
  author={Dagenais, Barth{\'e}l{\'e}my and Robillard, Martin P},
  journal={ACM Transactions on Software Engineering and Methodology (TOSEM)},
  volume={20},
  number={4},
  pages={1--35},
  year={2011},
  publisher={ACM New York, NY, USA}
}

@inproceedings{shi2022precise,
  title={Precise (un) affected version analysis for web vulnerabilities},
  author={Shi, Youkun and Zhang, Yuan and Luo, Tianhan and Mao, Xiangyu and Yang, Min},
  booktitle={Proceedings of the 37th IEEE/ACM International Conference on Automated Software Engineering},
  pages={1--13},
  year={2022}
}

@inproceedings{nguyen2013reliability,
  title={The (un) reliability of nvd vulnerable versions data: An empirical experiment on google chrome vulnerabilities},
  author={Nguyen, Viet Hung and Massacci, Fabio},
  booktitle={Proceedings of the 8th ACM SIGSAC symposium on Information, computer and communications security},
  pages={493--498},
  year={2013}
}

@article{sliwerski2005changes,
  title={When do changes induce fixes?},
  author={{\'S}liwerski, Jacek and Zimmermann, Thomas and Zeller, Andreas},
  journal={ACM sigsoft software engineering notes},
  volume={30},
  number={4},
  pages={1--5},
  year={2005},
  publisher={ACM New York, NY, USA}
}

@article{nguyen2016automatic,
  title={An automatic method for assessing the versions affected by a vulnerability},
  author={Nguyen, Viet Hung and Dashevskyi, Stanislav and Massacci, Fabio},
  journal={Empirical Software Engineering},
  volume={21},
  number={6},
  pages={2268--2297},
  year={2016},
  publisher={Springer}
}

@inproceedings{li2016vulpecker,
  title={Vulpecker: an automated vulnerability detection system based on code similarity analysis},
  author={Li, Zhen and Zou, Deqing and Xu, Shouhuai and Jin, Hai and Qi, Hanchao and Hu, Jie},
  booktitle={Proceedings of the 32nd annual conference on computer security applications},
  pages={201--213},
  year={2016}
}

@article{li2021sysevr,
  title={Sysevr: A framework for using deep learning to detect software vulnerabilities},
  author={Li, Zhen and Zou, Deqing and Xu, Shouhuai and Jin, Hai and Zhu, Yawei and Chen, Zhaoxuan},
  journal={IEEE Transactions on Dependable and Secure Computing},
  volume={19},
  number={4},
  pages={2244--2258},
  year={2021},
  publisher={IEEE}
}

@article{li2018vuldeepecker,
  title={Vuldeepecker: A deep learning-based system for vulnerability detection},
  author={Li, Zhen and Zou, Deqing and Xu, Shouhuai and Ou, Xinyu and Jin, Hai and Wang, Sujuan and Deng, Zhijun and Zhong, Yuyi},
  journal={arXiv preprint arXiv:1801.01681},
  year={2018}
}

@article{cheng2025vercation,
  title={VERCATION: Precise Vulnerable Open-source Software Version Identification based on Static Analysis and LLM},
  author={Cheng, Yiran and Zhang, Ting and Shar, Lwin Khin and Yang, Shouguo and Dong, Chaopeng and Lo, David and Lv, Shichao and Shi, Zhiqiang and Sun, Limin},
  journal={IEEE Transactions on Software Engineering},
  year={2025},
  publisher={IEEE}
}

@inproceedings{kim2017vuddy,
  title={Vuddy: A scalable approach for vulnerable code clone discovery},
  author={Kim, Seulbae and Woo, Seunghoon and Lee, Heejo and Oh, Hakjoo},
  booktitle={2017 IEEE symposium on security and privacy (SP)},
  pages={595--614},
  year={2017},
  organization={IEEE}
}

@inproceedings{xiao2020mvp,
  title={$\{$MVP$\}$: Detecting vulnerabilities using $\{$Patch-Enhanced$\}$ vulnerability signatures},
  author={Xiao, Yang and Chen, Bihuan and Yu, Chendong and Xu, Zhengzi and Yuan, Zimu and Li, Feng and Liu, Binghong and Liu, Yang and Huo, Wei and Zou, Wei and others},
  booktitle={29th USENIX Security Symposium (USENIX Security 20)},
  pages={1165--1182},
  year={2020}
}

@inproceedings{woo2022movery,
  title={$\{$MOVERY$\}$: A precise approach for modified vulnerable code clone discovery from modified $\{$Open-Source$\}$ software components},
  author={Woo, Seunghoon and Hong, Hyunji and Choi, Eunjin and Lee, Heejo},
  booktitle={31st USENIX Security Symposium (USENIX Security 22)},
  pages={3037--3053},
  year={2022}
}

@inproceedings{woo2023v1scan,
  title={$\{$V1SCAN$\}$: Discovering 1-day vulnerabilities in reused $\{$C/C++$\}$ open-source software components using code classification techniques},
  author={Woo, Seunghoon and Choi, Eunjin and Lee, Heejo and Oh, Hakjoo},
  booktitle={32nd USENIX Security Symposium (USENIX Security 23)},
  pages={6541--6556},
  year={2023}
}

@inproceedings{woo2021v0finder,
  title={$\{$V0Finder$\}$: Discovering the correct origin of publicly reported software vulnerabilities},
  author={Woo, Seunghoon and Lee, Dongwook and Park, Sunghan and Lee, Heejo and Dietrich, Sven},
  booktitle={30th USENIX Security Symposium (USENIX Security 21)},
  pages={3041--3058},
  year={2021}
}

@inproceedings{perl2015vccfinder,
  title={Vccfinder: Finding potential vulnerabilities in open-source projects to assist code audits},
  author={Perl, Henning and Dechand, Sergej and Smith, Matthew and Arp, Daniel and Yamaguchi, Fabian and Rieck, Konrad and Fahl, Sascha and Acar, Yasemin},
  booktitle={Proceedings of the 22nd ACM SIGSAC conference on computer and communications security},
  pages={426--437},
  year={2015}
}

\end{document}